\documentclass[aps,twocolumn,superscriptaddress]{revtex4-1}

\usepackage{amsmath,amssymb,amsfonts,color,graphicx,tabularx}

\usepackage[unicode=true,colorlinks=true]{hyperref}

\hypersetup{linkcolor=blue,citecolor=blue,urlcolor=blue}

\begin{document}

\title{Generalized Calogero-Sutherland models for Kondo physics in the Luttinger liquid}

\author{Hua-Chen Zhang}
\affiliation{Institut f\"ur Theoretische Physik, Technische Universit\"at Dresden, 01062 Dresden, Germany}
\affiliation{Beijing National Laboratory for Condensed Matter Physics \& Institute of Physics, Chinese Academy of Sciences, Beijing 100190, China}
\affiliation{School of Physical Sciences, University of Chinese Academy of Sciences, Beijing 100049, China}

\author{Ying-Hai Wu}
\email{yinghaiwu88@hust.edu.cn}
\affiliation{School of Physics and Wuhan National High Magnetic Field Center, Huazhong University of Science and Technology, Wuhan 430074, China}

\author{Hong-Hao Tu}
\email{hong-hao.tu@tu-dresden.de}
\affiliation{Institut f\"ur Theoretische Physik, Technische Universit\"at Dresden, 01062 Dresden, Germany}

\begin{abstract}
We propose a series of models with inverse-square interactions, for which the ground states have an exact closed form, that describe one localized spin-$1/2$ Kondo impurity coupled to Luttinger liquids of itinerant bosons or fermions in the continuum or on a lattice. The model Hamiltonians contain two parts: a two-component Calogero-Sutherland model with open boundary condition in the continuum or a long-range $t \textendash J$-type model on the lattice, and long-range Kondo couplings between the impurity and the particles. The wave functions of the ground states in a certain basis have Jastrow product forms with powers depending on the particle statistics. It is shown that the wave functions can be expressed as chiral correlators of certain conformal field theories, which reveals that the particles are fractionalized to fully screen the Kondo impurity. This insight would be very useful in future studies on the very difficult problem of Kondo physics with interacting baths.
\end{abstract}

\maketitle

\date{\today}

{\em Introduction.} --- One fundamental principle of quantum mechanics is that the Hilbert space of a composite system is the tensor product of the Hilbert spaces of its individual components. This beautiful structure brings about the exponential growth of the dimension of the Hilbert space, which becomes a formidable barrier as the number of constituents increases. Apart from the cases without interaction or with weak interactions, there is no universally valid and efficient approach for quantum many-body systems. For strongly correlated systems, one has to accept the fact that exact solutions cannot be found in most cases. Nevertheless, exactly solvable models have been constructed and analyzed shortly after the quantum mechanical revolution. The first example is the one-dimensional Heisenberg chain solved by Bethe~\cite{bethe1931} using an Ansatz that was later named after him. This method has been vastly generalized and provides deep insight into various aspects of mathematics and physics~\cite{korepin1993,gomez1996,essler2005}. In general, exactly solvable models could be appreciated from the aesthetic perspective, help us to understand certain universal properties of physical systems, and used to test some approaches that provide approximate solutions to many-body problems.

The present paper takes inspiration from the Calogero-Sutherland models~\cite{calogero1971,sutherland1971a,sutherland1971b,sutherland1971c,sutherland1972,ha1992,sutherland2004,kuramoto2009}. One representative of these models describes particles on a circle that interact via an inverse-sine-square potential. The ground state assumes a simple product form, which facilitates the exact computation of the excited states and certain correlation functions. The original papers already noticed intricate connections with random matrix theory and orthogonal polynomials (which has already been proposed in mathematical literature by Jack)~\cite{sutherland2004}. The discovery of the fractional quantum Hall (FQH) effect~\cite{tsui1982} led to the Laughlin wave function~\cite{laughlin1983} and subsequent investigations of fractional statistics in two-dimensional topological order~\cite{nayak2008}. The similarity between Calogero-Sutherland and Laughlin wave functions motivated extensive studies of exclusion statistics in ideal anyon gases~\cite{haldane1991,wu1994,murthy1994,wu1995,schoutens1997,wu2001}. In this Letter, we propose a series of models that inherits the spirit of the Calogero-Sutherland models and extends into the realm of quantum impurity physics.

The seminal works of Anderson~\cite{anderson1961} and Kondo~\cite{kondo1964} laid the foundation of quantum impurity systems, which have been studied in great detail due to their experimental relevance and theoretical perplexity. The prototypical Kondo model consists of a localized magnetic moment and free fermions coupled by the spin exchange interaction. Its simplicity is highly deceptive because a complete understanding requires knowledge of the concept of renormalization and the machinery of the numerical renormalization group (NRG)~\cite{wilson1975,krishna1980a,bulla2008}. While the original problem is defined in three dimensions, an equivalent one-dimensional model can be constructed owing to the non-interacting nature of the itinerant fermions. This mapping not only underlies the success of the NRG calculations, but also provides us some cases that can be solved exactly using the Bethe ansatz~\cite{andrei1980,wiegmann1980,andrei1983,tsvelik1983}.

\begin{figure}
\includegraphics[width=\columnwidth]{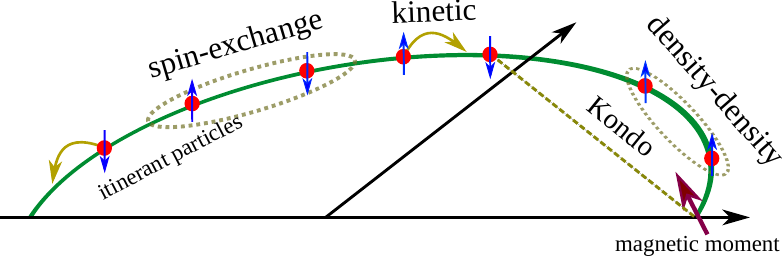}
\caption{Schematics of the model. One magnetic moment is placed at the right end of a semi-cicle populated by itinerant particles. The localized spin and itinerant particle are coupled by inverse-square Kondo terms. The itinerant particles interact with each other via density-density and spin-exchange interactions.}
\label{fig:model}
\end{figure}

For quantum impurity systems, considering interactions in the bath makes the problem much harder, and which makes this subject much less understood~\cite{lee1992,furusaki1994,schiller1995,froejdh1995,fendley1996,wang1997,wang2001,rylands2016,rylands2017,rylands2018}. The approach based on many-body wave functions~\cite{beduerftig1999,pasnoori2020} can often provide crucial insights into the properties of such systems, and the present work takes a significant step along this direction. As illustrated in Fig.~\ref{fig:model}, our models contain a spin (the magnetic impurity) fixed at one end of a semi-circle and itinerant bosons or fermions within the semi-circle. The systems may be defined in the continuum or on some lattice sites, much as the relation between the Calogero-Sutherland model and the Haldane-Shastry model~\cite{haldane1988,shastry1988}. The impurity and the particles are coupled by long-range Kondo terms, whilst the particles move on the semi-circle and have mutual long-range density-density and spin-spin interactions. The bath realizes a Luttinger liquid~\cite{tomonaga1950,luttinger1963,mattis1965,haldane1981}, whose interplay with Kondo physics is poorly understood. The ground-state wave functions have product forms that are reminiscent of the multi-component Halperin FQH states~\cite{halperin1983} and can be formulated as chiral correlators of certain conformal field theories (CFTs)~\cite{cirac2010,nielsen2011,tu2015,basumallick2016,hackenbroich2017,tu2019,wu2019}.

{\em Continuum models.} --- For the continuum case, we have $2M-1$ spin-$1/2$ itinerant particles on the semi-circle with unity radius (see Fig.~\ref{fig:model}). The angular positions of these particles are denoted as $x_{j} \in (0,\pi],~j = 1, 2, \ldots, 2M-1$, whose projections on the horizontal axis are $z_{j} = \cos{x_j} \in [-1,1)$. We always impose the ordering $x_{1} < x_{2} < \cdots < x_{2M-1}$ on these variables. The impurity is a spin-$1/2$ magnetic moment fixed at the right end with coordinate $x_{0} \equiv 0$. The spin indices of the impurity and the itinerant particles are denoted as $\sigma_{0}$ and $\sigma_{j}$, respectively.

We consider the first-quantized many-body wave function
\begin{align}
\label{eq:wave-function-continuum}
    \Psi(\{z_{j},\sigma_{j}\}) =~ &\delta_{\sum_{j=0}^{2M-1}\delta_{\sigma_{j},\uparrow} =\sum_{j=0}^{2M-1}\delta_{\sigma_{j},\downarrow} = M} \nonumber \\
    \times & \left[ \prod_{j=1}^{2M-1} e^{\frac{i\pi}{2}\mathrm{sgn}(\sigma_{0}-\sigma_{j})} (1-z_{j})^{\delta_{\sigma_{0},\sigma_{j}}} \right]
    \Psi_{\textrm{bulk}}
\end{align}
with
\begin{equation}
\label{eq:wave-function-continuum-bulk}
    \Psi_{\textrm{bulk}} = \prod_{1 \leq j < k \leq 2M-1} e^{\frac{i\pi}{2}\mathrm{sgn}(\sigma_{j}-\sigma_{k})} (z_{j} - z_{k})^{\lambda + \delta_{\sigma_{j},\sigma_{k}}}.
\end{equation}
The Kronecker $\delta$ in Eq.~\eqref{eq:wave-function-continuum} imposes constraints on the state of the localized spin and the number of itinerant particles: There are $M-1$ spin-up (spin-down) particles [and the other $M$ particles have spin-down (spin-up)] when the impurity assumes the up (down) state. The value of $\mathrm{sgn}(\sigma_{j}-\sigma_{k})$ is $+1$ if $\sigma_{j}=\uparrow,\sigma_{k}=\downarrow$, $-1$ if $\sigma_{j}=\downarrow,\sigma_{k}=\uparrow$, and $0$ if $\sigma_{j}=\sigma_{k}$. A real parameter $\lambda \geq 0$ appears in the power of the Jastrow products. The bulk part in Eq.~\eqref{eq:wave-function-continuum-bulk} is very similar to the Halperin $(\lambda + 1, \lambda + 1, \lambda)$ FQH state~\cite{halperin1983}. However, the statistics of the itinerant particles has not been specified yet. It can be shown that the wave function is a global spin singlet (see the Supplemental Material for the proof), in conformity with the ``screening'' scenario of Kondo physics.

Remarkably, there exist a number of operators that annihilate the model wave function in Eq.~\eqref{eq:wave-function-continuum}, which facilitates the construction of an exact parent Hamiltonian (see the Supplemental Material for the details). First, it is proved that the ``current'' operators,
\begin{equation}
    \Lambda^{\alpha}_{j} = \sum_{k=0 \atop (k \neq j)}^{2M-1} \frac{\sin x_{j}}{z_{j} - z_{k}} \left[ i \left( \mathbf{S}_{j} \times \mathbf{S}_{k} \right)^{\alpha} + S^{\alpha}_{k} \right]
\end{equation}
with $\alpha = x, y, z$ and $j = 1, 2, \ldots, 2M-1$, annihilate $\Psi$. Furthermore, this wave function satisfies Knizhnik-Zamolodchikov-type differential equations~\cite{knizhnik1984}, $\Omega_{j} \Psi = 0$, where
\begin{align}
    \Omega_{j} = \frac{\partial}{\partial x_{j}} + \sin x_{j} \Bigg[ &\frac{2}{3} \sum_{k=0 \atop (k \neq j)}^{2M-1} \frac{1}{z_{j} - z_{k}} \left( \mathbf{S}_{j} \cdot \mathbf{S}_{k} + \frac{3}{4} \right) \nonumber \\
    &+ \lambda \sum_{k=1 \atop (k \neq j)}^{2M-1} \frac{1}{z_{j} - z_{k}} \Bigg].
\end{align}
This, together with the singlet nature of $\Psi$, implies that the latter is an exact ground state of the positive semi-definite Hamiltonian $\sum_{j} ( \Omega_{j}^{\dagger} \Omega_{j} + \mu_{1} \sum_{\alpha} {\Lambda^{\alpha}_{j}}^{\dagger} \Lambda^{\alpha}_{j} ) + \mu_{2} \mathbf{S}^{2}$ with $\mathbf{S}$ the total spin operator and arbitrary $\mu_{1}, \mu_{2} \geq 0$. By taking $\mu_{1} = \frac{4}{9} (\lambda + \frac{1}{3})$, $\mu_{2} = \frac{\lambda}{3} (4M - 3) + \frac{1}{3} (2M - 1)$ and a suitable constant shift (given in the Supplemental Material), the parent Hamiltonian admits the intriguing structure $H = H_{0} + H_{\textrm{B}} + H_{\textrm{I}}$, where
\begin{equation}
\label{eq:hamiltonian-continuum-free}
    H_{0} = -\sum_{j} \frac{\partial^{2}}{\partial x_{j}^{2}}
\end{equation}
is the kinetic term of the itinerant particles,
\begin{align}
\label{eq:hamiltonian-continuum-bulk}
    H_{\textrm{B}} =~&\lambda(2\lambda + 1) \sum_{j<k} \left[ \frac{1}{\left( d(x_{j},x_{k}) \right)^{2}} + \frac{1}{\left( \bar{d}(x_{j},x_{k}) \right)^{2}} \right] \nonumber \\
    &+ 4\lambda \sum_{j<k} \left[ \frac{1}{\left( d(x_{j},x_{k}) \right)^{2}} + \frac{1}{\left( \bar{d}(x_{j},x_{k}) \right)^{2}} \right] \mathbf{S}_{j} \cdot \mathbf{S}_{k}
\end{align}
includes inverse-square density-density and spin-exchange interactions between the itinerant particles, with $d(x_{j},x_{k}) = 2 \sin\frac{x_{j}-x_{k}}{2}$ and $\bar{d}(x_{j},x_{k}) = 2 \sin\frac{x_{j}+x_{k}}{2}$, and
\begin{align}
\label{eq:hamiltonian-continuum-impurity}
    H_{\textrm{I}} = ~&(2\lambda + 1) \sum_{j} \frac{1}{\left( d(0,x_{j}) \right)^{2}} \nonumber \\
    &+ \frac{4}{3}(2\lambda + 1)\sum_{j} \frac{1}{\left( d(0,x_{j}) \right)^{2}}~\mathbf{S}_{0} \cdot \mathbf{S}_{j}
\end{align}
denotes a boundary potential as well as the Kondo coupling. The summations in Eqs.~\eqref{eq:hamiltonian-continuum-free},~\eqref{eq:hamiltonian-continuum-bulk} and~\eqref{eq:hamiltonian-continuum-impurity} run over the labels $j, k = 1, 2, \ldots, 2M-1$ of the itinerant particles. If the $H_{\textrm{I}}$ term is excluded, the Hamiltonian reduces to that of the $D_{N}$-type two-component Calogero-Sutherland model with an open boundary~\cite{yamamoto1995,serban1997,basumallick2011}. The presence of the impurity makes the problem more difficult, and we have not been able to find additional closed-form eigenstates other than Eq.~\eqref{eq:wave-function-continuum}.

{\em Lattice models.} ---  To convert the continuum models to lattice versions, we introduce lattice sites at angular positions $x_{j}=\pi(j-1/2)/N$ ($j=1,\ldots,N$) such that the itinerant particles only hop between these sites (the impurity is still fixed at $x_{0}=0$). One important revision is that the parameter $\lambda$ is now restricted to positive integers. The itinerant particles are fermions (hardcore bosons) when $\lambda$ is even (odd), as one can see from the symmetry or anti-symmetry of Eq.~\eqref{eq:wave-function-continuum} under particle exchange.

The counterpart of Eq.~\eqref{eq:wave-function-continuum} on the lattice is (up to an unimportant overall factor)
\begin{equation}
\label{eq:wave-function-lattice}
    \vert \widetilde{\Psi} \rangle = \vert \Uparrow \rangle \otimes \vert \Psi^{\downarrow} \rangle - \vert \Downarrow \rangle \otimes \vert \Psi^{\uparrow} \rangle,
\end{equation}
where $\vert\Uparrow\rangle$ ($\vert\Downarrow\rangle$) denote the spin-up (spin-down) state of the impurity and the itinerant particles are described by
\begin{widetext}
\begin{align}
\label{eq:wave-function-lattice-component1}
    \vert \Psi^{\downarrow} \rangle = \sum_{u_{1}>\ldots>u_{M-1}} \sum_{v_{1}>\ldots>v_{M}} &\left( \prod_{k=1}^{M-1} (1 - u_{k}) \right) \left( \prod_{1 \leq k < l \leq M-1} (u_{k} - u_{l})^{\lambda+1} \prod_{1 \leq m < n \leq M} (v_{m} - v_{n})^{\lambda+1} \right) \nonumber \\
    &\times \left( \prod_{k=1}^{M-1} \prod_{m=1}^{M} (u_{k} - v_{m})^{\lambda} \right) c_{u_{1},\uparrow}^{\dagger} \cdots c_{u_{M-1},\uparrow}^{\dagger} c_{v_{1},\downarrow}^{\dagger} \cdots c_{v_{M},\downarrow}^{\dagger} \vert 0 \rangle,
\end{align}
and
\begin{align}
\label{eq:wave-function-lattice-component2}
    \vert \Psi^{\uparrow} \rangle = (-1)^{M-1} \sum_{u_{1}>\ldots>u_{M}} \sum_{v_{1}>\ldots>v_{M-1}} &\left( \prod_{m=1}^{M-1} (1 - v_{m}) \right) \left( \prod_{1 \leq k < l \leq M} (u_{k} - u_{l})^{\lambda+1} \prod_{1 \leq m < n \leq M-1} (v_{m} - v_{n})^{\lambda+1} \right) \nonumber \\
    &\times \left( \prod_{k=1}^{M} \prod_{m=1}^{M-1} (u_{k} - v_{m})^{\lambda} \right) c_{u_{1},\uparrow}^{\dagger} \cdots c_{u_{M},\uparrow}^{\dagger} c_{v_{1},\downarrow}^{\dagger} \cdots c_{v_{M-1},\downarrow}^{\dagger} \vert 0 \rangle.
\end{align}
\end{widetext}
The variables $u_{k}$ ($v_{m}$) are drawn from the set of projected coordinates $\{ z_{\mathrm{1}}, \ldots, z_{N} \}$ of the lattice sites. $|0\rangle$ is the vacuum for the itinerant particles and $c_{z_{j},\sigma}^{\dagger}$ creates a fermion (hardcore boson) with spin $\sigma$ at the site $j$ when $\lambda$ is even (odd).

A parent Hamiltonian on the lattice can be found if $(2\lambda + 1)(M-1) \leq N$. This constraint means that the total power of $\{ u_{k} \}$ and $\{ v_{m} \}$ in the wave function does not exceed $N$. By defining particle number operators $n_{j} \equiv \sum_{\sigma} c_{z_{j},\sigma}^{\dagger}c_{z_{j},\sigma}$, the interaction terms in Eqs.~\eqref{eq:hamiltonian-continuum-bulk} and~\eqref{eq:hamiltonian-continuum-impurity} can be transcribed to their lattice counterparts $\widetilde{H}_{\textrm{B}}$ and $\widetilde{H}_{\textrm{I}}$. The operators $n_{j} n_{k}$ and $n_j$ are appended to the density-density interaction and boundary potential terms (after the existing coefficients), respectively. Moreover, the summations over $j$ and $k$ are with respect to lattice sites rather than the particle indices. The lattice hopping representation of the kinetic term in Eq.~\eqref{eq:hamiltonian-continuum-free} is less obvious. Making use of the fact that the single-particle eigenstates on the semi-circle are given by discrete Chebyshev polynomials, we prove that
\begin{align}
\label{eq:hamiltonian-lattice-free}
    \widetilde{H}_{0} = &\sum_{j=1}^{N} \left[ \frac{1}{3}N^{2} + \frac{1}{6} - \frac{1}{2(1-z_{j}^2)} \right] n_{j} \nonumber \\
    &+  \sum_{j \neq k}^{N} \sum_{\sigma} \left[ \frac{2(-1)^{j-k}}{\left( d(x_{j},x_{k}) \right)^{2}} - \frac{2(-1)^{j-k}}{\left( \bar{d}(x_{j},x_{k}) \right)^{2}} \right] c_{z_{k},\sigma}^{\dagger}c_{z_{j},\sigma}
\end{align}
is a suitable hopping term (see the Supplemental Material for the proof). The lattice wave function in Eq.~\eqref{eq:wave-function-lattice} is an exact eigenstate of the Hamiltonian $\widetilde{H} = \mathcal{P} (\widetilde{H}_{0} + \widetilde{H}_{\textrm{B}} + \widetilde{H}_{\textrm{I}}) \mathcal{P}$ with the {\it same} eigenvalue as in the continuum, where $\mathcal{P}$ is the projector that removes double occupancy of itinerant bosons or fermions. Hence, $\widetilde{H}$ has the form of a $t \textendash J$-type Hamiltonian coupled to a Kondo impurity, in which the hopping and interaction strengths decay with distance in an inverse-square manner.

\begin{figure}
\includegraphics[width=\columnwidth]{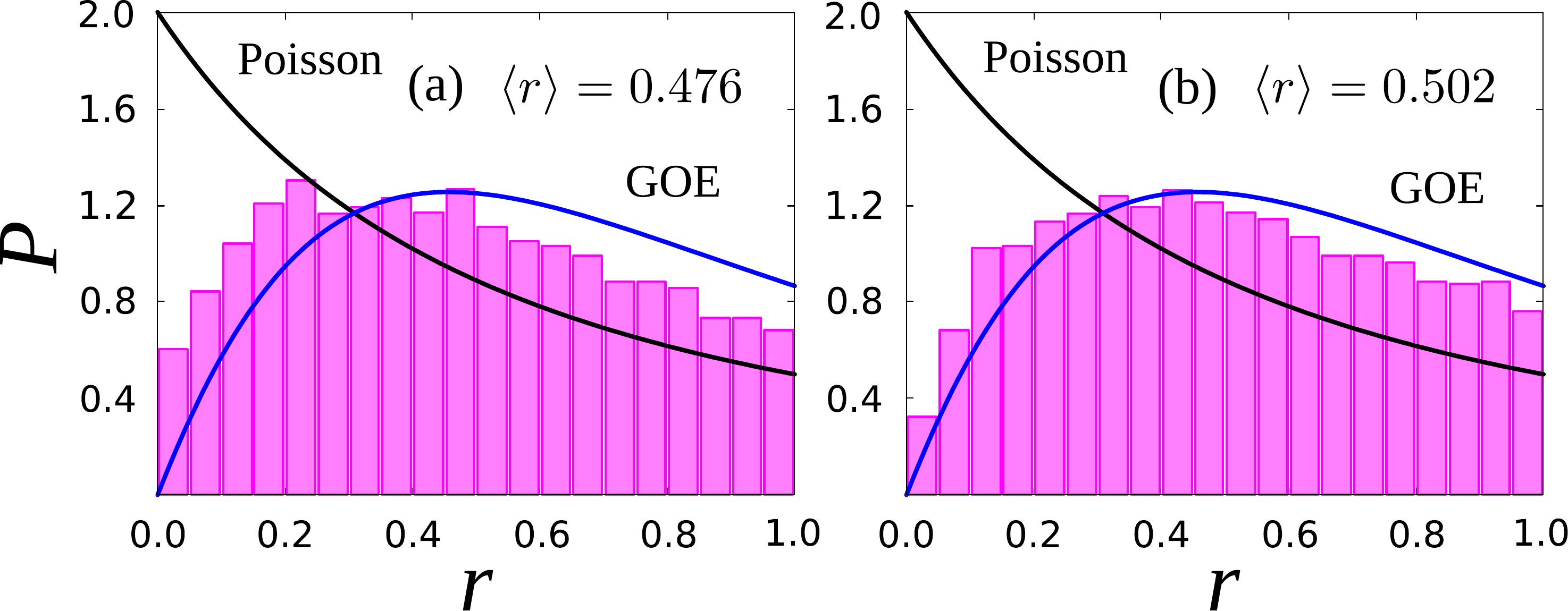}
\caption{Energy level statistics in the subspace with $\mathbf{S}^{2}=0$ and $S^{z}=0$. The probability density of the ratio $r_{m}$ is represented by the bars. The Poisson and GOE results are drawn for reference. (a) $\lambda=1$, $N=14$, and $M=3$. (b) $\lambda=2$, $N=14$, and $M=3$.}
\label{fig:level}
\end{figure}

{\em Numerical results.} ---~The intimate connections between the lattice model and its continuum counterpart strongly suggest that the wave function in Eq.~\eqref{eq:wave-function-lattice} is also an exact ground state of the corresponding lattice Hamiltonian. To confirm this conjecture, we have performed exact diagonalization of the lattice Hamiltonian for $\lambda=1,2,3,4$ with up to $N=16$ sites and various choices of $M$. The wave function in Eq.~\eqref{eq:wave-function-lattice} has also been explicitly constructed. In all cases, the ground-state energy agrees with the analytical prediction and the ground state has unit overlap with Eq.~\eqref{eq:wave-function-lattice}.

Exact solutions for certain Kondo problems in terms of the Bethe Ansatz reveal the integrability of such models. In the absence of the impurity, the lattice Hamiltonian with $\lambda=1$ describes the SU(3) Haldane-Shastry model with open boundary condition, which is integrable and has a remarkable twisted Yangian symmetry~\cite{tu2015}. It is thus natural to ask whether or not our models are integrable. To this end, we have studied the energy level spacing statistics in several cases~\cite{poilblanc1993,Atas2013}. The good quantum numbers are the total spin $\mathbf{S}^{2}$ and its $z$ component $S^{z}$. All eigenvalues in the subspace with $\mathbf{S}^{2}=0$ and $S^{z}=0$ are computed and sorted in ascending order. The level spacing $\delta_{m}=E_{m+1}-E_{m}$ and the ratio $r_{m}=\min(\delta_{m},\delta_{m+1})/\max(\delta_{m},\delta_{m+1})$ are defined. The probability density $P(r)$ of $r_{m}$ is expected to obey the Poissonian form with mean value $\langle r \rangle=0.386$ for integrable models and the Gaussian orthogonal ensemble (GOE) with $\langle r \rangle=0.536$ for other real Hamiltonians. However, exotic behaviors have been reported in certain Haldane-Shastry type models~\cite{finkel2005,Barba2008a}. It seems that our models are more likely to be in the GOE class as shown in Fig.~\ref{fig:level}. Moreover, numerical calculations of the out-of-time-ordered correlator also suggest that these models are non-integrable (see the Supplemental Material for more details). One heuristic argument for the non-integrability is that many eigenvalues of the Hamiltonian are not rational numbers despite the existence of a small subset that is rational. This is in sharp contrast to the SU(3) Haldane-Shastry model with open boundary condition, whose eigenvalues are all rational upon suitable normalization. While the lattice models are likely non-integrable, the possibility that the continuum models are integrable cannot be ruled out yet.

{\em Conformal field theory formulation.} --- It was pointed out in the works of Affleck and Ludwig that boundary CFT can be used to investigate Kondo physics~\cite{affleck1991a,affleck1991c}. In our model, there is an interesting connection between the many-body wave function and the boundary CFT that describes the low-energy physics. It turns out that the wave function in Eq.~\eqref{eq:wave-function-lattice} can be expressed as the following CFT correlators (see the Supplemental Material for more details):
\begin{align}
\label{eq:IDMPS-component1}
    &\langle \Uparrow; u_{1}, \ldots, u_{M-1}, v_{1}, \ldots, v_{M} \vert \widetilde{\Psi} \rangle \nonumber \\
    &= \langle \mathcal{O}_{\textrm{bg}} A_{0}^{\uparrow}(1) A^{\uparrow}(u_{1}) \cdots A^{\uparrow}(u_{M-1}) A^{\downarrow}(v_{1}) \cdots A^{\downarrow}(v_{M}) \rangle,
\end{align}
and
\begin{align}
\label{eq:IDMPS-component2}
    &\langle \Downarrow; u_{1}, \ldots, u_{M}, v_{1}, \ldots, v_{M-1} \vert \widetilde{\Psi} \rangle \nonumber \\
    &= \langle \mathcal{O}_{\textrm{bg}} A_{0}^{\downarrow}(1) A^{\uparrow}(u_{1}) \cdots A^{\uparrow}(u_{M}) A^{\downarrow}(v_{1}) \cdots A^{\downarrow}(v_{M-1}) \rangle,
\end{align}
where $\mathcal{O}_{\textrm{bg}}$ is a neutralizing background charge, and the vertex operators
\begin{align}
    A_{0}^{\uparrow/\downarrow}(1) &= \kappa_{\uparrow/\downarrow} :e^{\pm \frac{i}{\sqrt{2}}\phi_{\textrm{s}}(1)} e^{\frac{i}{\sqrt{2(2\lambda + 1)}}\phi_{\textrm{c}}(1)}:, \\
    A^{\uparrow}(u_{k}) &= \kappa_{\uparrow} :e^{\frac{i}{\sqrt{2}}\phi_{\textrm{s}}(u_{k})} e^{i\sqrt{\frac{2\lambda + 1}{2}}\phi_{\textrm{c}}(u_{k})}:, \\
    A^{\downarrow}(v_{m}) &= \kappa_{\downarrow} :e^{-\frac{i}{\sqrt{2}}\phi_{\textrm{s}}(v_{m})} e^{i\sqrt{\frac{2\lambda + 1}{2}}\phi_{\textrm{c}}(v_{m})}:
\end{align}
are constructed from a free two-component bosonic field ($\phi_{\textrm{s}}$ and $\phi_{\textrm{c}}$), and $:\cdots:$ denotes normal ordering~\cite{francesco1997}. The Klein factors $\kappa_{\uparrow/\downarrow}$ satisfying the Majorana commutation relation ($\kappa_{\uparrow}^{2} = \kappa_{\downarrow}^{2} = 1$ and $\{ \kappa_{\uparrow},\kappa_{\downarrow} \} = 0$) are introduced to generate correct statistics for the particles.

The formulation in terms of CFT correlators uncovers valuable physical insight into the Kondo physics in a Luttinger liquid. A well-known fact about Luttinger liquids is that the spin and charge degrees of freedom decouple with each other and are described separately by the two components $\phi_{\textrm{s}}$ and $\phi_{\textrm{c}}$ of a free boson CFT. The vertex operators $A^{\uparrow}$ and $A^{\downarrow}$ constructed from the CFT fields have conformal dimension $\frac{\lambda+1}{2}$. It deviates from that of a free fermion, which is $1/2$, for all cases with $\lambda>0$. This marks the breakdown of the Fermi liquid description that is applicable when the itinerant particles in the bath have no interactions. On the other hand, the vertex operator $A_{0}^{\uparrow/\downarrow}$ that represents the Kondo impurity has conformal dimension $\frac{\lambda+1}{2(2\lambda+1)}$. In fact, the form of $A_{0}^{\uparrow/\downarrow}$ is precisely the same as that of the quasihole operator in the $(\lambda + 1, \lambda + 1, \lambda)$ Halperin FQH state~\cite{moore1991}, which has spin $1/2$ and charge $-1/(2\lambda + 1)$. Intuitively, one itinerant particle is ``split'' into $2\lambda + 1$ pieces such that the magnetic moment can be screened properly. This phenomenon of~\emph{fractionalization} is a hallmark of the strong interaction in the Luttinger liquid. Interestingly, a similar notion of fractionalization was proposed in multichannel Kondo problems, where bosonization was utilized to map the system to a Luttinger liquid coupled to an impurity~\cite{emery1992}.

{\em Conclusion and discussion.} --- To summarize, we have proposed a series of quantum impurity models, for which the ground states have an exact closed form, in one-dimensional space with open boundaries. The bulk is populated by spin-$1/2$ particles and one edge hosts a spin-$1/2$ impurity, which are coupled by inverse-square long-range density-density, spin-exchange, and Kondo interactions. The models involve a tunable parameter $\lambda$ that assumes a non-negative real value. If $\lambda$ is a positive integer, the lattice version of the model can be obtained, where the itinerant particles are fermions (hard-core bosons) for even (odd) $\lambda$. In the ground state, the spin of the itinerant particles and that of the impurity form a singlet. The low-energy theory of the interacting bulk is a Luttinger liquid, so our models capture the spin-$1/2$ Kondo physics in a Luttinger liquid. Further insight into the nature of the ground state is revealed by the fact that the wave function can be expressed as CFT correlators. The operator on the impurity site serves as the ``boundary condition changing operator'' for the bulk Luttinger liquid. It is still not proved that the model wave functions are unique ground states of the corresponding Hamiltonians. Although numerical results suggest that they are indeed unique in the lattice models, an analytical proof is not easy. It may be fruitful to exploit the fact that nodes are absent in the wave functions. Another open question is to understand analytically the (non-)integrability of our models, for which the exchange operator formalism may offer an interesting perspective~\cite{polychronakos1992,fowler1993,polychronakos1993,ghosh1998,yang2022}. Moreover, the connection with bilayer Halperin FQH states calls for a more detailed study of exclusion statistics in our models. There should also be natural generalizations to some analogs of non-Abelian multi-component FQH states~\cite{ardonne1999,barkeshli2010}. Finally, it is worth noting that Jastrow-type wave functions were also found in exactly solvable Kondo-lattice-like models~\cite{wang1995,wang1996}. We hope that this work would motivate further investigations on Kondo physics in Luttinger liquids and more general interacting baths.

{\em Acknowledgments.} --- We thank Meng Cheng, Jan von Delft, and Jing Yang for helpful discussions. This work was supported by the Deutsche Forschungsgemeinschaft (DFG) through project A06 of SFB 1143 (project-id 247310070), the National Key Research and Development Project of China (Grant No.~2017YFA0302901), the National Natural Science Foundation of China (Grants No.~11888101, No.~11874095, and No.~12174130), the Strategic Priority Research Program of Chinese Academy of Sciences (Grant No.~XDB33000000), and startup grant of HUST. H.C.Z. acknowledges a scholarship from the CAS-DAAD Joint Fellowship Programme.

\bibliography{CSKL-v2}

\onecolumngrid

\setcounter{figure}{0}
\setcounter{equation}{0}
\renewcommand\thefigure{S\arabic{figure}}
\renewcommand\theequation{S\arabic{equation}}

\vspace{20pt}
\begin{center}
\large \textbf{Supplemental Material}
\end{center}

In this Supplemental Material, we provide the proof of the singlet property of our model wave function, present the construction of the exact parent Hamiltonian as well as an alternative proof of the eigenvalue equation, establish the equivalence between our model wave function and certain conformal field theory (CFT) correlators, display the low-energy spectrum of the lattice model, and illustrate the numerical results on the out-of-time-ordered correlator in the ground state.

\section{Singlet property}
\label{sec:singlet-property}

In this section, we prove that the model wave function defined in Eqs.~\eqref{eq:wave-function-continuum} and~\eqref{eq:wave-function-continuum-bulk} is a global spin-singlet. By definition, this amounts to proving $\mathbf{S}^{2} \Psi = 0$, where $\mathbf{S} \equiv \sum_{j=0}^{2M-1} \mathbf{S}_{j}$ is the total spin operator. It is easy to see that
\begin{equation}
    \mathbf{S}^{2} = \Big( \sum_{j^{\prime}} S_{j^{\prime}}^{-} \Big) \Big( \sum_{j} S_{j}^{+} \Big) + \Big( \sum_{j} S_{j}^{z} \Big) + \Big( \sum_{j} S_{j}^{z} \Big)^{2},
\end{equation}
where $S_{j}^{+} \equiv S_{j}^{x} + iS_{j}^{y}$ and $S_{j}^{-} \equiv S_{j}^{x} - iS_{j}^{y}$ are spin ladder operators. Due to the existence of the Kronecker $\delta$ factor in~\eqref{eq:wave-function-continuum}, it is obvious that $(\sum_{j=0}^{2M-1} S_{j}^{z})\Psi = 0$. Therefore, it suffices to prove $(\sum_{j=0}^{2M-1} S_{j}^{+})\Psi = 0$. For the sake of notational simplicity, we denote the wave function in~\eqref{eq:wave-function-continuum} apart from the Kronecker $\delta$ factor as $\Psi_{0}$, namely
\begin{equation}
    \Psi(\{z_{j},\sigma_{j}\}) = \delta_{\sum_{j=0}^{2M-1}\delta_{\sigma_{j},\uparrow} =\sum_{j=0}^{2M-1}\delta_{\sigma_{j},\downarrow} = M} \Psi_{0}(\{z_{j},\sigma_{j}\}).
\end{equation}
When acting on $\Psi(\{z_{j},\sigma_{j}\})$ with $\sigma_{j} = \downarrow$, $S_{j}^{+}$ flips $\sigma_{j}$ to $\uparrow$; if $\sigma_{j} = \uparrow$ in $\Psi(\{z_{j},\sigma_{j}\})$, the latter will be annihilated by the action of $S_{j}^{+}$. Thus, the wave function after the action of $\sum_{j=0}^{2M-1} S_{j}^{+}$ has the form
\begin{equation}
    \Big( \sum_{j=0}^{2M-1} S_{j}^{+} \Big) \Psi(\{z_{j},\sigma_{j}\}) = \delta_{\sum_{j=0}^{2M-1}\delta_{\sigma_{j},\uparrow} = M+1,~\sum_{j=0}^{2M-1}\delta_{\sigma_{j},\downarrow} = M-1} \Psi^{\prime}_{0}(\{z_{j},\sigma_{j}\}).
\end{equation}
In fact, from the expression~\eqref{eq:wave-function-continuum} it is straightforward to obtain that
\begin{equation}
\label{eq:singlet-proof}
    \Psi^{\prime}_{0} = e^{\frac{i\pi}{2}} (-1)^{M} \left[ \delta_{\sigma_{0},\uparrow} \prod_{k=1}^{2M-1} (1 - z_{k})^{\delta_{\sigma_{k},\downarrow} - \delta_{\sigma_{k},\uparrow}} - \sum_{j=1}^{2M-1} \delta_{\sigma_{j},\uparrow} (1 - z_{j})^{\delta_{\sigma_{0},\downarrow} - \delta_{\sigma_{0},\uparrow}} \prod_{k(\neq j)}^{2M-1} (z_{j} - z_{k})^{\delta_{\sigma_{k},\downarrow} - \delta_{\sigma_{k},\uparrow}} \right] \Psi_{0}.
\end{equation}
If $\sigma_{0} = \uparrow$ in $\Psi^{\prime}_{0}(\{z_{j},\sigma_{j}\})$, we find
\begin{align}
\label{eq:singlet-proof-continued1}
    &\text{the expression inside the bracket in~\eqref{eq:singlet-proof}} = \prod_{k=1}^{2M-1} (1 - z_{k})^{\delta_{\sigma_{k},\downarrow} - \delta_{\sigma_{k},\uparrow}} - \sum_{j=1}^{2M-1} \delta_{\sigma_{j},\uparrow} \frac{1}{1 - z_{j}} \prod_{k(\neq j)}^{2M-1} (z_{j} - z_{k})^{\delta_{\sigma_{k},\downarrow} - \delta_{\sigma_{k},\uparrow}} \nonumber \\
    =&~\frac{\prod_{n=1}^{M-1}(1 - v_{n})}{\prod_{m=1}^{M}(1 - u_{m})} - \sum_{j=1}^{M} \frac{1}{1 - u_{j}} \frac{\prod_{n=1}^{M-1}(u_{j} - v_{n})}{\prod_{m(\neq j)}^{M}(u_{j} - u_{m})},
\end{align}
in obtaining the last identity, we have renamed the coordinates of the spin-up particles as $u_{1} > u_{2} > \cdots > u_{M}$ and those of the spin-down particles as $v_{1} > v_{2} > \cdots > v_{M-1}$. As we shall see, the right-hand-side of~\eqref{eq:singlet-proof-continued1} can readily be proven to be $0$ by using techniques from complex analysis. To this end, let us define an auxiliary function
\begin{equation}
    f(z) = \frac{1}{1-z} \frac{\prod_{n=1}^{M-1}(z - v_{n})}{\prod_{m=1}^{M}(z - u_{m})},
\end{equation}
where $z$ is a complex variable. $f(z)$ has simple poles at $z = 1$ and $z = u_{j},~j = 1, 2, \ldots, M$, where the residues of $f(z)$ are
\begin{equation}
    \mathrm{Res}(f,1) = -\frac{\prod_{n=1}^{M-1}(1 - v_{n})}{\prod_{m=1}^{M}(1 - u_{m})}
\end{equation}
and
\begin{equation}
    \mathrm{Res}(f,u_{j}) = \frac{1}{1-u_{j}} \frac{\prod_{n=1}^{M-1}(u_{j} - v_{n})}{\prod_{m(\neq j)}^{M}(u_{j} - u_{m})},
\end{equation}
respectively. Moreover, by Laurent expanding $f(z)$ around infinity, one finds $\mathrm{Res}(f,\infty) = 0$. According to Cauchy's theorem in complex analysis, one has
\begin{equation}
    \mathrm{Res}(f,1) + \sum_{j=1}^{M} \mathrm{Res}(f,u_{j}) = 0,
\end{equation}
which is precisely the statement that the right-hand-side of~\eqref{eq:singlet-proof-continued1} vanishes. On the other hand, if $\sigma_{0} = \downarrow$ in $\Psi^{\prime}_{0}(\{z_{j},\sigma_{j}\})$,
\begin{equation}
\label{eq:singlet-proof-continued2}
    \text{the expression inside the bracket in~\eqref{eq:singlet-proof}} = -\sum_{j=1}^{M+1} (1 - u_{j}) \frac{\prod_{n=1}^{M-2}(u_{j} - v_{n})}{\prod_{m(\neq j)}^{M+1}(u_{j} - u_{m})}.
\end{equation}
Defining the auxiliary function
\begin{equation}
    g(z) = (1 - z) \frac{\prod_{n=1}^{M-2}(z - v_{n})}{\prod_{m=1}^{M+1}(z - u_{m})},
\end{equation}
one finds
\begin{equation}
    \mathrm{Res}(g,u_{j}) = (1 - u_{j}) \frac{\prod_{n=1}^{M-2}(u_{j} - v_{n})}{\prod_{m(\neq j)}^{M+1}(u_{j} - u_{m})}
\end{equation}
and $\mathrm{Res}(g,\infty) = 0$. According to the same reasoning as above, the right-hand-side of~\eqref{eq:singlet-proof-continued2} vanishes as well. This concludes the proof of the singlet property of wave function in~\eqref{eq:wave-function-continuum}. We note that the summation technique based on complex analysis introduced above will also be extensively used in Sec.~\ref{sec:parent-hamiltonian}.

\section{Parent Hamiltonian}
\label{sec:parent-hamiltonian}

\subsection{Construction based on annihilators}

In this subsection, we present the construction of the exact parent Hamiltonian for the model wave function in~\eqref{eq:wave-function-continuum}; the construction is based on the existence of operators that annihilate this wave function.

To begin with, let us act the operator $(S_{j}^{+}S_{k}^{-} + S_{j}^{-}S_{k}^{+})$ with $j = 1, \ldots, (2M-1)$, $k = 0, 1, \ldots, (2M-1)$ and $j \neq k$ on $\Psi$. As the effect of acting with this operator is to exchange the spin coordinates of particles (if $k \neq 0$; or those of the impurity and a particle if $k=0$) $j$ and $k$ if $\sigma_{j} \neq \sigma_{k}$ and annihilate $\Psi$ if $\sigma_{j} = \sigma_{k}$, one finds
\begin{equation}
    (S_{j}^{+}S_{k}^{-} + S_{j}^{-}S_{k}^{+}) \Psi = -(1 - \delta_{\sigma_{j},\sigma_{k}}) \prod_{l=0 \atop (l \neq j,k)}^{2M-1} (z_{j} - z_{l})^{\delta_{\sigma_{k},\sigma_{l}} - \delta_{\sigma_{j},\sigma_{l}}} (z_{k} - z_{l})^{\delta_{\sigma_{j},\sigma_{l}} - \delta_{\sigma_{k},\sigma_{l}}} \Psi.
\end{equation}
By relabelling the spatial coordinates of the particles (or the impurity) with spin coordinate equal (resp. opposite) to $\sigma_{j}$ as $u_{1} > u_{2} > \cdots > u_{M}$ (resp. $v_{1} > v_{2} > \cdots > v_{M}$) and observing the notation $z_{j} \equiv u_{x_{j}}, z_{k} \equiv v_{y_{k}}$, the above equation becomes
\begin{equation}
    (S_{j}^{+}S_{k}^{-} + S_{j}^{-}S_{k}^{+}) \Psi = -(1 - \delta_{\sigma_{j},\sigma_{k}}) \prod_{x=1 \atop (x \neq x_{j})}^{M} \left( \frac{v_{y_{k}} - u_{x}}{u_{x_{j}} - u_{x}} \right) \prod_{y=1 \atop (y \neq y_{k})}^{M} \left( \frac{u_{x_{j}} - v_{y}}{v_{y_{k}} - v_{y}} \right) \Psi,
\end{equation}
and hence
\begin{equation}
    \sum_{k=0 \atop (k \neq j)}^{2M-1} \frac{1}{z_{j} - z_{k}} (S_{j}^{+}S_{k}^{-} + S_{j}^{-}S_{k}^{+}) \Psi = - \sum_{y_{k} = 1}^{M} \frac{1}{u_{x_{j}} - v_{y_{k}}} \prod_{x=1 \atop (x \neq x_{j})}^{M} \left( \frac{v_{y_{k}} - u_{x}}{u_{x_{j}} - u_{x}} \right) \prod_{y=1 \atop (y \neq y_{k})}^{M} \left( \frac{u_{x_{j}} - v_{y}}{v_{y_{k}} - v_{y}} \right) \Psi
\end{equation}
Using the technique based on complex analysis introduced in Sec.~\ref{sec:singlet-property}, the summation on the right-hand-side of the above equation can be carried out as follows. We introduce an auxiliary function
\begin{equation}
    F(z) = \frac{1}{(z - u_{x_{j}})^{2}} \prod_{x=1 \atop (x \neq x_{j})}^{M} \left( \frac{z - u_{x}}{u_{x_{j}} - u_{x}} \right) \prod_{y=1}^{M} \left( \frac{u_{x_{j}} - v_{y}}{z - v_{y}} \right)
\end{equation}
with complex variable $z$. $F(z)$ has $M$ simple poles at $z = v_{y_{k}}, y_{k} = 1, 2, \ldots, M$, where the residues are
\begin{equation}
    \mathrm{Res}(F, v_{y_{k}}) = \frac{1}{u_{x_{j}} - v_{y_{k}}} \prod_{x=1 \atop (x \neq x_{j})}^{M} \left( \frac{v_{y_{k}} - u_{x}}{u_{x_{j}} - u_{x}} \right) \prod_{y=1 \atop (y \neq y_{k})}^{M} \left( \frac{u_{x_{j}} - v_{y}}{v_{y_{k}} - v_{y}} \right).
\end{equation}
Moreover, $F(z)$ has a second-order pole at $z = u_{x_{j}}$, where the residue is
\begin{align}
    \mathrm{Res}(F, u_{x_{j}}) &= \frac{\mathrm{d}}{\mathrm{d}z} \Bigg[ \prod_{x=1 \atop (x \neq x_{j})}^{M} \left( \frac{z - u_{x}}{u_{x_{j}} - u_{x}} \right) \prod_{y=1}^{M} \left( \frac{u_{x_{j}} - v_{y}}{z - v_{y}} \right) \Bigg]\Bigg\vert_{z = u_{x_{j}}} \nonumber \\
    &= \sum_{x=1 \atop (x \neq x_{j})}^{M} \frac{1}{u_{x_{j}} - u_{x}} - \sum_{y=1}^{M} \frac{1}{u_{x_{j}} - v_{y}}.
\end{align}
As $\mathrm{Res}(F,\infty) = 0$, one obtains by Cauchy's theorem that
\begin{align}
    \sum_{k=0 \atop (k \neq j)}^{2M-1} \frac{1}{z_{j} - z_{k}} (S_{j}^{+}S_{k}^{-} + S_{j}^{-}S_{k}^{+}) \Psi &= - \sum_{y_{k} = 1}^{M} \mathrm{Res}(F, v_{y_{k}}) \Psi = \mathrm{Res}(F, u_{x_{j}}) \Psi \nonumber \\
    &= 2 \sum_{k=0 \atop (k \neq j)}^{2M-1} \frac{1}{z_{j} - z_{k}} \delta_{\sigma_{j},\sigma_{k}} \Psi - \sum_{k=0 \atop (k \neq j)}^{2M-1} \frac{1}{z_{j} - z_{k}} \Psi,
\end{align}
or, equivalently,
\begin{equation}
\label{eq:annihilator1}
    \sum_{k=0 \atop (k \neq j)}^{2M-1} \frac{1}{z_{j} - z_{k}} \left( S_{j}^{x}S_{k}^{x} + S_{j}^{y}S_{k}^{y} - 2S_{j}^{z}S_{k}^{z} \right) \Psi = 0.
\end{equation}
A similar line of reasoning yields
\begin{equation}
    \sum_{k=0 \atop (k \neq j)}^{2M-1} \frac{1}{z_{j} - z_{k}} \left[ i\left( S_{j}^{x}S_{k}^{y} - S_{j}^{y}S_{k}^{x} \right) + S_{k}^{z} \right] \Psi = 0;
\end{equation}
we define the annihilators
\begin{equation}
    \Lambda^{z}_{j} = \sin x_{j} \cdot \sum_{k=0 \atop (k \neq j)}^{2M-1} \frac{1}{z_{j} - z_{k}} \left[ i\left( S_{j}^{x}S_{k}^{y} - S_{j}^{y}S_{k}^{x} \right) + S_{k}^{z} \right]
\end{equation}
with $j = 1, 2, \ldots, (2M-1)$. The singlet property of $\Psi$ (see Sec.~\ref{sec:singlet-property}) implies that the global spin rotations of $\Lambda^{z}_{j}$, i.e., $\Lambda^{x}_{j}$ and $\Lambda^{y}_{j}$, also annihilate the same wave function. These annihilators are collectively expressed as
\begin{equation}
\label{eq:current-operators}
    \Lambda^{\alpha}_{j} = \sin x_{j} \cdot \sum_{k=0 \atop (k \neq j)}^{2M-1} \frac{1}{z_{j} - z_{k}} \left( i\varepsilon_{\alpha\beta\gamma}S_{j}^{\beta}S_{k}^{\gamma} + S_{k}^{\alpha} \right),
\end{equation}
where $\alpha, \beta, \gamma = x, y, z$ are the indices for spin components, $\varepsilon_{\alpha\beta\gamma}$ is the totally antisymmetric Levi-Civita symbol with $\varepsilon_{xyz} \equiv +1$ and repeated spin-component indices are summed over (the same convention will be used in the following). We note in passing that these annihilators can also be derived by CFT techniques, namely, by the decoupling of the $\mathrm{SU}(2)_{1}$ null field~\cite{nielsen2011}.

On the other hand, by directly computing the partial derivative of $\Psi$ with respect to $x_{j}$, $j = 1, 2, \ldots, (2M-1)$, one obtains the Knizhnik-Zamolodchikov (KZ)-type differential equations~\cite{knizhnik1984},
\begin{equation}
\label{eq:KZ}
    \frac{\partial \Psi}{\partial x_{j}} + \sin x_{j} \cdot \left[ \sum_{k=0 \atop (k \neq j)}^{2M-1} \frac{1}{z_{j} - z_{k}} \left( 2S_{j}^{z}S_{k}^{z} + \frac{1}{2} \right) + \lambda \sum_{k=1 \atop (k \neq j)}^{2M-1} \frac{1}{z_{j} - z_{k}} \right] \Psi = 0.
\end{equation}
These equations are not $\mathrm{SU}(2)$-symmetric; combining~\eqref{eq:KZ} with~\eqref{eq:annihilator1} yields the $\mathrm{SU}(2)$-symmetric version of the KZ-type equations, $\Omega_{j} \Psi = 0$, where
\begin{equation}
    \Omega_{j} = \frac{\partial}{\partial x_{j}} + \sin x_{j} \left[ \frac{2}{3} \sum_{k=0 \atop (k \neq j)}^{2M-1} \frac{1}{z_{j} - z_{k}} \left( \mathbf{S}_{j} \cdot \mathbf{S}_{k} + \frac{3}{4} \right) + \lambda \sum_{k=1 \atop (k \neq j)}^{2M-1} \frac{1}{z_{j} - z_{k}} \right].
\end{equation}

With the sets of annihilators, $\{ \Lambda^{\alpha}_{j} \}$ and $\{ \Omega_{j} \}$, as well as the singlet nature of $\Psi$, one can construct a quite general Hermitian Hamiltonian
\begin{equation}
    H^{\prime} = \sum_{j=1}^{2M-1} ( \Omega_{j}^{\dagger} \Omega_{j} + \mu_{1} {\Lambda^{\alpha}_{j}}^{\dagger} \Lambda^{\alpha}_{j} ) + \mu_{2} \mathbf{S}^{2} + E,
\end{equation}
where $\mu_{1}$, $\mu_{2}$ and $E$ are arbitrary real constants. The model wave function $\Psi$ is an exact eigenstate of this Hamiltonian,
\begin{equation}
\label{eq:eigenvalue-equation}
    H^{\prime} \Psi = E \Psi;
\end{equation}
in particular, if $\mu_{1}, \mu_{2} \geq 0$, $\Psi$ is a~\emph{ground} state of $H^{\prime}$ since $(H^{\prime} - E)$ is now positive semi-definite.

It is tedious but not so difficult to derive
\begin{align}
\label{eq:expression-part1}
    \sum_{j=1}^{2M-1} {\Lambda^{\alpha}_{j}}^{\dagger} \Lambda^{\alpha}_{j} =~&\frac{9}{4} \sum_{j=1}^{2M-1} \frac{1}{1 - z_{j}} + 3 \sum_{j=1}^{2M-1} \frac{1}{1 - z_{j}} \mathbf{S}_{0} \cdot \mathbf{S}_{j} + \frac{9}{8} \sum_{j \neq k}^{2M-1} \frac{1 - z_{j}^{2}}{(z_{j} - z_{k})^{2}} + \frac{3}{2} \sum_{j \neq k}^{2M-1} \frac{1 - z_{j}^{2}}{(z_{j} - z_{k})^{2}} \mathbf{S}_{j} \cdot \mathbf{S}_{k} \nonumber \\
    &- 3 \sum_{j \neq k}^{2M-1} \frac{1 + z_{j}}{z_{j} - z_{k}} \mathbf{S}_{0} \cdot \mathbf{S}_{k} + \frac{3}{2} \sum_{j \neq k \neq l}^{2M-1} \frac{1 - z_{j}^{2}}{(z_{j} - z_{k})(z_{j} - z_{l})} \mathbf{S}_{k} \cdot \mathbf{S}_{l} - \frac{3}{2} \mathbf{S}_{0} \cdot \sum_{j=1}^{2M-1} \mathbf{S}_{j} - \frac{9}{8} (2M - 1)
\end{align}
and
\begin{align}
\label{eq:expression-part2}
    \sum_{j=1}^{2M-1} \Omega_{j}^{\dagger} \Omega_{j} = &-\sum_{j=1}^{2M-1} \frac{\partial^{2}}{\partial x_{j}^{2}} + \frac{1}{6} \sum_{j=1}^{2M-1} \frac{1}{1 - z_{j}} + \frac{2}{9} \sum_{j=1}^{2M-1} \frac{1}{1 - z_{j}} \mathbf{S}_{0} \cdot \mathbf{S}_{j} \nonumber \\
    &+ (\lambda^{2} - \frac{1}{6}) \sum_{j \neq k}^{2M-1} \frac{1 - z_{j}^{2}}{(z_{j} - z_{k})^{2}} + (\frac{4}{3} \lambda - \frac{2}{9}) \sum_{j \neq k}^{2M-1} \frac{1 - z_{j}^{2}}{(z_{j} - z_{k})^{2}} \mathbf{S}_{j} \cdot \mathbf{S}_{k} \nonumber \\
    &+ \frac{4}{3} (\lambda + \frac{1}{3}) \sum_{j \neq k}^{2M-1} \frac{1 + z_{j}}{z_{j} - z_{k}} \mathbf{S}_{0} \cdot \mathbf{S}_{k} - \frac{2}{3} (\lambda + \frac{1}{3}) \sum_{j \neq k \neq l}^{2M-1} \frac{1 - z_{j}^{2}}{(z_{j} - z_{k})(z_{j} - z_{l})} \mathbf{S}_{k} \cdot \mathbf{S}_{l} \nonumber \\
    &- \left[ \frac{4}{3} \lambda (2M - 2) + \frac{2}{3} (2M - \frac{4}{3}) \right] \mathbf{S}_{0} \cdot \sum_{j=1}^{2M-1} \mathbf{S}_{j} - \left[ \frac{2}{3} \lambda (2M - 3) + \frac{1}{3} (2M - 1) \right] \sum_{j \neq k}^{2M-1} \mathbf{S}_{j} \cdot \mathbf{S}_{k} \nonumber \\
    &- \frac{2}{3} \lambda^{2} (M - 1) (2M - 1) (2M - 3) - \frac{4}{3} \lambda M (M - 1) (2M - 1) - \frac{1}{6} (2M - 1) (2M^{2} + M - 1);
\end{align}
in the derivation, we have used the $\mathrm{SU}(2)$ algebra $S^{\alpha}S^{\beta} = \delta_{\alpha\beta}/4 + i\varepsilon_{\alpha\beta\gamma}S^{\gamma}/2$, the identity $\varepsilon_{\alpha\beta\gamma} \varepsilon_{\alpha^{\prime}\beta^{\prime}\gamma} = \delta_{\alpha\alpha^{\prime}} \delta_{\beta\beta^{\prime}} - \delta_{\alpha\beta^{\prime}} \delta_{\beta\alpha^{\prime}}$ satisfied by the Levi-Civita symbol and the summation identities
\begin{equation}
\label{eq:auxiliary-identity-0-2}
    \sum_{j \neq k}^{2M-1} \frac{z_{j}}{z_{j} - z_{k}} = \frac{1}{2} \sum_{j \neq k}^{2M-1} \Bigg( \frac{z_{j}}{z_{j} - z_{k}} + \frac{z_{k}}{z_{k} - z_{j}} \Bigg) = \frac{1}{2} \sum_{j \neq k}^{2M-1} 1 = (M - 1) (2M - 1),
\end{equation}
\begin{align}
\label{eq:auxiliary-identity-0-1}
    &\sum_{j \neq k \neq l}^{2M-1} \frac{1 - z_{j}^{2}}{(z_{j} - z_{k})(z_{j} - z_{l})} = \frac{1}{3} \sum_{j \neq k \neq l}^{2M-1} \Bigg[ \frac{1 - z_{j}^{2}}{(z_{j} - z_{k})(z_{j} - z_{l})} + \frac{1 - z_{k}^{2}}{(z_{k} - z_{l})(z_{k} - z_{j})} + \frac{1 - z_{l}^{2}}{(z_{l} - z_{j})(z_{l} - z_{k})} \Bigg] \nonumber \\
    =& -\frac{1}{3} \sum_{j \neq k \neq l}^{2M-1} 1 = -\frac{1}{3} (2M - 1) (2M - 2) (2M - 3).
\end{align}
By inspecting the explicit expressions in~\eqref{eq:expression-part1} and~\eqref{eq:expression-part2}, it is straightforward to see that if we choose $\mu_{1} = \frac{4}{9} (\lambda + \frac{1}{3})$, $\mu_{2} = \frac{\lambda}{3} (4M - 3) + \frac{1}{3} (2M - 1)$ and
\begin{equation}
\label{eq:eigenenergy}
    E = \frac{\lambda^{2}}{3}(M-1)(2M-1)(4M-3) + \frac{\lambda}{3}M(M-1)(8M-7) + \frac{1}{3}M(M-1)(2M-1),
\end{equation}
the Hamiltonian $H^{\prime} = H$ is nothing but the parent Hamiltonian introduced in the Main Text. We remark that our construction here is quite reminiscent of that in Refs.~\cite{shastry1992,bernevig2001}, in which the Haldane-Shastry model was recast into a factorized, or ``frustration-free'', form.

\subsection{Alternative proof
of the eigenvalue equation}

In this subsection, we provide an alternative proof
of the eigenvalue equation~\eqref{eq:eigenvalue-equation} by directly acting the Hamiltonian $H = H_{0} + H_{\textrm{B}} + H_{\textrm{I}}$ on the wave function $\Psi$. First, let us separate out the terms of two-body potential from $H_{\textrm{B}}$ and $H_{\textrm{I}}$ by writing $H_{\textrm{B}} = H_{\textrm{B1}} + H_{\textrm{B2}}$, $H_{\textrm{I}} = H_{\textrm{I1}} + H_{\textrm{I2}}$, where
\begin{equation}
    H_{\textrm{B1}} = \lambda(2\lambda + 1) \sum_{j<k} \left[ \frac{1}{\left( d(x_{j},x_{k}) \right)^{2}} + \frac{1}{\left( \bar{d}(x_{j},x_{k}) \right)^{2}} \right], \qquad H_{\textrm{I1}} = (2\lambda + 1) \sum_{j} \frac{1}{\left( d(0,x_{j}) \right)^{2}}
\end{equation}
are the two-body potentials;
\begin{equation}
    H_{\textrm{B2}} = 4\lambda \sum_{j<k} \left[ \frac{1}{\left( d(x_{j},x_{k}) \right)^{2}} + \frac{1}{\left( \bar{d}(x_{j},x_{k}) \right)^{2}} \right] \mathbf{S}_{j} \cdot \mathbf{S}_{k}
\end{equation}
and
\begin{equation}
    H_{\textrm{I2}} = \frac{4}{3}(2\lambda + 1)\sum_{j} \frac{1}{\left( d(0,x_{j}) \right)^{2}}~\mathbf{S}_{0} \cdot \mathbf{S}_{j}
\end{equation}
are the terms of spin-exchange interaction in the bulk and Kondo interaction, respectively. Let us consider each of these terms separately.

\subsubsection{The kinetic term}

By direct calculation, one finds
\begin{align}
    H_{0} \Psi = & -\sum_{j=1}^{2M-1} \frac{\partial^{2}}{\partial x_{j}^{2}} \Psi \nonumber \\
    = & \Bigg[ -\lambda(\lambda - 1) \sum_{j \neq k}^{2M-1} \frac{1 - z_{j}z_{k}}{(z_{j} - z_{k})^{2}} + \lambda^{2} \sum_{j \neq k}^{2M-1} \frac{z_{j}}{z_{j} - z_{k}} - \lambda^{2} \sum_{j \neq k \neq l}^{2M-1} \frac{1 - z_{j}^{2}}{(z_{j} - z_{k})(z_{j} - z_{l})} \nonumber \\
    &~~ + 2 \lambda \sum_{j \neq k}^{2M-1} \frac{1 + z_{j}}{z_{j} - z_{k}} \delta_{\sigma_{0},\sigma_{j}} - 2 \lambda \sum_{j \neq k}^{2M-1} \frac{1 - z_{j}^{2}}{(z_{j} - z_{k})^{2}} \delta_{\sigma_{j},\sigma_{k}} - 2 \lambda \sum_{j \neq k \neq l}^{2M-1} \frac{1 - z_{j}^{2}}{(z_{j} - z_{k})(z_{j} - z_{l})} \delta_{\sigma_{j},\sigma_{k}} \nonumber \\
    &~~ - \sum_{j=1}^{2M-1} \frac{z_{j}}{1 - z_{j}} \delta_{\sigma_{0},\sigma_{j}} + \sum_{j \neq k}^{2M-1} \frac{z_{j}}{z_{j} - z_{k}} \delta_{\sigma_{j},\sigma_{k}} + 2 \sum_{j \neq k}^{2M-1} \frac{1 + z_{j}}{z_{j} - z_{k}} \delta_{\sigma_{0},\sigma_{j}} \delta_{\sigma_{0},\sigma_{k}} \nonumber \\
    &~~ - \sum_{j \neq k \neq l}^{2M-1} \frac{1 - z_{j}^{2}}{(z_{j} - z_{k})(z_{j} - z_{l})} \delta_{\sigma_{j},\sigma_{k}} \delta_{\sigma_{j},\sigma_{l}} \Bigg] \Psi.
\end{align}
In fact, some of the terms on the right-hand-side of the above expression can be proven to be constants; these are~\eqref{eq:auxiliary-identity-0-2},~\eqref{eq:auxiliary-identity-0-1} and
\begin{equation}
\label{eq:auxiliary-identity-0}
    \sum_{j \neq k}^{2M-1} \frac{z_{j}}{z_{j} - z_{k}} \delta_{\sigma_{j},\sigma_{k}} = \frac{1}{2} \sum_{j \neq k}^{2M-1} \delta_{\sigma_{j},\sigma_{k}} = (M - 1)^{2},
\end{equation}
\begin{equation}
\label{eq:auxiliary-identity-1}
    \sum_{j \neq k}^{2M-1} \frac{1 + z_{j}}{z_{j} - z_{k}} \delta_{\sigma_{0},\sigma_{j}} \delta_{\sigma_{0},\sigma_{k}} = \frac{1}{2} \sum_{j \neq k}^{2M-1} \delta_{\sigma_{0},\sigma_{j}} \delta_{\sigma_{0},\sigma_{k}} = \frac{1}{2} (M - 1) (M - 2),
\end{equation}
\begin{equation}
\label{eq:auxiliary-identity-2}
    \sum_{j \neq k \neq l}^{2M-1} \frac{1 - z_{j}^{2}}{(z_{j} - z_{k})(z_{j} - z_{l})} \delta_{\sigma_{j},\sigma_{k}} \delta_{\sigma_{j},\sigma_{l}} = -\frac{1}{3} \sum_{j \neq k \neq l}^{2M-1} \delta_{\sigma_{j},\sigma_{k}} \delta_{\sigma_{j},\sigma_{l}} = -\frac{1}{3} (M - 1) (M - 2) (2M - 3).
\end{equation}
Thus, we have
\begin{align}
\label{eq:H0-action-result}
    H_{0} \Psi =  \Bigg[ &-\lambda(\lambda - 1) \sum_{j \neq k}^{2M-1} \frac{1 - z_{j}z_{k}}{(z_{j} - z_{k})^{2}} + 2 \lambda \sum_{j \neq k}^{2M-1} \frac{1 + z_{j}}{z_{j} - z_{k}} \delta_{\sigma_{0},\sigma_{j}} \nonumber \\
    &- 2 \lambda \sum_{j \neq k}^{2M-1} \frac{1 - z_{j}^{2}}{(z_{j} - z_{k})^{2}} \delta_{\sigma_{j},\sigma_{k}} - 2 \lambda \sum_{j \neq k \neq l}^{2M-1} \frac{1 - z_{j}^{2}}{(z_{j} - z_{k})(z_{j} - z_{l})} \delta_{\sigma_{j},\sigma_{k}} \nonumber \\
    &- \sum_{j=1}^{2M-1} \frac{z_{j}}{1 - z_{j}} \delta_{\sigma_{0},\sigma_{j}} + \frac{\lambda^{2}}{3} (M - 1) (2M - 1) (4M - 3) + \frac{1}{3} (M + 1) (M - 1) (2M - 3) \Bigg] \Psi.
\end{align}

\subsubsection{The term of spin-exchange interaction in the bulk}

By definition, $d(x_{j},x_{k}) = 2~\vert \mathrm{sin}\frac{x_{j}-x_{k}}{2} \vert$ and $\bar{d}(x_{j},x_{k}) = 2~\mathrm{sin}\frac{x_{j}+x_{k}}{2}$, we have
\begin{equation}
    \frac{1}{\left( d(x_{j},x_{k}) \right)^{2}} + \frac{1}{\left( \bar{d}(x_{j},x_{k}) \right)^{2}} = \frac{1 - z_{j}z_{k}}{(z_{j} - z_{k})^{2}}.
\end{equation}
Moreover, we notice that the spin exchange interaction can be rewritten in terms of permutation operators:
\begin{equation}
    \mathbf{S}_{j} \cdot \mathbf{S}_{k} = \frac{1}{4}~(2\mathrm{P}_{jk} - 1),
\end{equation}
where $\mathrm{P}_{jk}$ is the permutation operator acting on $\mathbb{C}^{2} \otimes \mathbb{C}^{2}$, the tensor product of the spin Hilbert spaces of particles $j$ and $k$: $\mathrm{P}_{jk}(\mathcal{V} \otimes \mathcal{W}) = \mathcal{W} \otimes \mathcal{V},~\forall~\mathcal{V}, \mathcal{W} \in \mathbb{C}^{2}$, where $\otimes$ denotes the Kronecker product between vectors. In other words, the action of $\mathrm{P}_{jk}$ amounts to exchanging the spins of particles $j$ and $k$. Thus, the term of spin-exchange interactions in the bulk can be rewritten as
\begin{equation}
    H_{\textrm{B2}} = \frac{1}{2} \lambda \sum_{j \neq k}^{2M-1} \frac{1 - z_{j}z_{k}}{(z_{j} - z_{k})^{2}}~(2\mathrm{P}_{jk} - 1).
\end{equation}
Acting it on the wave function in~\eqref{eq:wave-function-continuum}, one finds
\begin{equation}
\label{eq:HB-action}
    H_{\textrm{B2}} \Psi = -\frac{1}{2} \lambda \sum_{j \neq k}^{2M-1} \frac{1 - z_{j}z_{k}}{(z_{j} - z_{k})^{2}} \Psi + \lambda \sum_{j \neq k}^{2M-1} \frac{1 - z_{j}z_{k}}{(z_{j} - z_{k})^{2}} \delta_{\sigma_{j},\sigma_{k}} \Psi + \lambda \sum_{j \neq k}^{2M-1} \frac{1 - z_{j}z_{k}}{(z_{j} - z_{k})^{2}} (1 - \delta_{\sigma_{j},\sigma_{k}}) \mathrm{P}_{jk} \Psi.
\end{equation}
The only non-trivial term on the right-hand-side of~\eqref{eq:HB-action} is the last one. Let us consider this term separately:
\begin{align}
    &\sum_{j \neq k}^{2M-1} \frac{1 - z_{j}z_{k}}{(z_{j} - z_{k})^{2}} (1 - \delta_{\sigma_{j},\sigma_{k}}) \mathrm{P}_{jk} \Psi \nonumber \\
    =& -\sum_{j \neq k}^{2M-1} \frac{1 - z_{j}z_{k}}{(z_{j} - z_{k})^{2}} (1 - \delta_{\sigma_{j},\sigma_{k}}) \left( \frac{1 - z_{k}}{1 - z_{j}} \right)^{\delta_{\sigma_{0},\sigma_{j}} - \delta_{\sigma_{0},\sigma_{k}}} \prod_{l(\neq j,k)}^{2M-1} \left( \frac{z_{l} - z_{k}}{z_{l} - z_{j}} \right)^{\delta_{\sigma_{j},\sigma_{l}} - \delta_{\sigma_{k},\sigma_{l}}} \Psi \nonumber \\
    =& -2 \sum_{j=1}^{M-1} \sum_{k=1}^{M} \frac{1 - u_{j}v_{k}}{(u_{j} - v_{k})^{2}} \frac{1 - v_{k}}{1 - u_{j}} \prod_{m(\neq j)}^{M-1} \left( \frac{u_{m} - v_{k}}{u_{m} - u_{j}} \right) \prod_{n(\neq k)}^{M} \left( \frac{v_{n} - u_{j}}{v_{n} - v_{k}} \right) \Psi,
\end{align}
in obtaining the last identity, we have renamed the coordinates of the particles with spin $\sigma_{0}$ as $u_{1} > u_{2} > \ldots > u_{M-1}$ and those with spin opposite to $\sigma_{0}$ as $v_{1} > v_{2} > \ldots > v_{M}$. Now let us carry out the summation on the right-hand-side of the above expression, which can be written as the sum of two parts:
\begin{align}
\label{eq:summation-to-be-evaluate}
    &\sum_{j=1}^{M-1} \sum_{k=1}^{M} \frac{1 - u_{j}v_{k}}{(u_{j} - v_{k})^{2}} \frac{1 - v_{k}}{1 - u_{j}} \prod_{m(\neq j)}^{M-1} \left( \frac{u_{m} - v_{k}}{u_{m} - u_{j}} \right) \prod_{n(\neq k)}^{M} \left( \frac{v_{n} - u_{j}}{v_{n} - v_{k}} \right) \nonumber \\
    =& \sum_{j=1}^{M-1} \sum_{k=1}^{M} \frac{1 - u_{j}v_{k}}{(1 - u_{j})(u_{j} - v_{k})} \prod_{m(\neq j)}^{M-1} \left( \frac{u_{m} - v_{k}}{u_{m} - u_{j}} \right) \prod_{n(\neq k)}^{M} \left( \frac{v_{n} - u_{j}}{v_{n} - v_{k}} \right) \nonumber \\
    &+ \sum_{j=1}^{M-1} \sum_{k=1}^{M} \frac{1 - u_{j}v_{k}}{(u_{j} - v_{k})^{2}} \prod_{m(\neq j)}^{M-1} \left( \frac{u_{m} - v_{k}}{u_{m} - u_{j}} \right) \prod_{n(\neq k)}^{M} \left( \frac{v_{n} - u_{j}}{v_{n} - v_{k}} \right).
\end{align}
The summations in both parts can again be done by applying the technique based on complex analysis introduced in Sec.~\ref{sec:singlet-property}. We illustrate this calculation using the first part as an example. To this end, we define a series of auxiliary functions
\begin{equation}
    f_{j}(z) = \frac{1 - u_{j}z}{(1 - u_{j})(u_{j} - z)^{2}} \prod_{m(\neq j)}^{M-1} \left( \frac{u_{m} - z}{u_{m} - u_{j}} \right) \prod_{n=1}^{M} \left( \frac{v_{n} - u_{j}}{v_{n} - z} \right),\quad j = 1, \ldots, M-1,
\end{equation}
where $z$ is the complex variable, $\{u_{m}\}$ and $\{v_{n}\}$ are parameters. Obviously, $f_{j}(z)$ has $M$ simple poles $z = v_{k}, k = 1, \ldots, M$ and one second-order pole $z = u_{j}$. At these poles, the residues of $f_{j}(z)$ are
\begin{equation}
    \mathrm{Res}(f_{j},v_{k}) = \frac{1 - u_{j}v_{k}}{(1 - u_{j})(u_{j} - v_{k})} \prod_{m(\neq j)}^{M-1} \left( \frac{u_{m} - v_{k}}{u_{m} - u_{j}} \right) \prod_{n(\neq k)}^{M} \left( \frac{v_{n} - u_{j}}{v_{n} - v_{k}} \right)
\end{equation}
and
\begin{align}
    \mathrm{Res}(f_{j},u_{j}) &= \frac{\mathrm{d}}{\mathrm{d}z} \Bigg[ \frac{1 - u_{j}z}{1 - u_{j}} \prod_{m(\neq j)}^{M-1} \left( \frac{u_{m} - z}{u_{m} - u_{j}} \right) \prod_{n=1}^{M} \left( \frac{v_{n} - u_{j}}{v_{n} - z} \right) \Bigg]\Bigg\vert_{z = u_{j}} \nonumber \\
    &= -\frac{u_{j}}{1 - u_{j}} + (1 + u_{j}) \left( \sum_{j^{\prime}(\neq j)}^{M-1} \frac{1}{u_{j} - u_{j^{\prime}}} - \sum_{k=1}^{M} \frac{1}{u_{j} - v_{k}}\right),
\end{align}
respectively. Moreover, by Laurent expanding $f_{j}(z)$ around infinity, one finds $\mathrm{Res}(f_{j},\infty) = 0$. According to Cauchy’s theorem in complex analysis, one has
\begin{equation}
    \mathrm{Res}(f_{j},u_{j}) + \sum_{k=1}^{M} \mathrm{Res}(f_{j},v_{k}) = 0;
\end{equation}
taking the summation over $j = 1, \ldots, M-1$ on both sides of the above equation, one immediately finds
\begin{equation}
\label{eq:part1}
    \sum_{j=1}^{M-1} \sum_{k=1}^{M} \frac{1 - u_{j}v_{k}}{(1 - u_{j})(u_{j} - v_{k})} \prod_{m(\neq j)}^{M-1} \left( \frac{u_{m} - v_{k}}{u_{m} - u_{j}} \right) \prod_{n(\neq k)}^{M} \left( \frac{v_{n} - u_{j}}{v_{n} - v_{k}} \right) = \sum_{j=1}^{M-1} \frac{u_{j}}{1 - u_{j}} - \sum_{j \neq j^{\prime}}^{M-1} \frac{1 + u_{j}}{u_{j} - u_{j^{\prime}}} + \sum_{j=1}^{M-1} \sum_{k=1}^{M} \frac{1 + u_{j}}{u_{j} - v_{k}}.
\end{equation}
The second part can be calculated in a similar manner, the result is
\begin{align}
\label{eq:part2}
    & \sum_{j=1}^{M-1} \sum_{k=1}^{M} \frac{1 - u_{j}v_{k}}{(u_{j} - v_{k})^{2}} \prod_{m(\neq j)}^{M-1} \left( \frac{u_{m} - v_{k}}{u_{m} - u_{j}} \right) \prod_{n(\neq k)}^{M} \left( \frac{v_{n} - u_{j}}{v_{n} - v_{k}} \right) \nonumber \\
    =& \sum_{j=1}^{M-1} \sum_{k=1}^{M} \frac{1 - u_{j}v_{k}}{(u_{j} - v_{k})^{2}} \left( 1 + \sum_{m(\neq j)}^{M-1} \frac{u_{j} - v_{k}}{u_{m} - u_{j}} - \sum_{n(\neq k)}^{M} \frac{u_{j} - v_{k}}{v_{n} - v_{k}} \right) + \frac{1}{3} M(M - 1)(M - 2).
\end{align}
Substituting~\eqref{eq:part1} and~\eqref{eq:part2} into the right-hand-side of~\eqref{eq:summation-to-be-evaluate}, one gets
\begin{align}
\label{eq:summation-evaluated}
    &\sum_{j=1}^{M-1} \sum_{k=1}^{M} \frac{1 - u_{j}v_{k}}{(u_{j} - v_{k})^{2}} \frac{1 - v_{k}}{1 - u_{j}} \prod_{m(\neq j)}^{M-1} \left( \frac{u_{m} - v_{k}}{u_{m} - u_{j}} \right) \prod_{n(\neq k)}^{M} \left( \frac{v_{n} - u_{j}}{v_{n} - v_{k}} \right) \nonumber \\
    =&~\frac{1}{2} \sum_{j \neq k}^{2M-1} \frac{1 - z_{j}z_{k}}{(z_{j} - z_{k})^{2}} (1 - \delta_{\sigma_{j},\sigma_{k}}) + \sum_{j \neq k}^{2M-1} \frac{1 + z_{j}}{z_{j} - z_{k}} \delta_{\sigma_{0},\sigma_{j}} + \sum_{j=1}^{2M-1} \frac{z_{j}}{1 - z_{j}} \delta_{\sigma_{0},\sigma_{j}} - 2 \sum_{j \neq k}^{2M-1} \frac{1 + z_{j}}{z_{j} - z_{k}} \delta_{\sigma_{0},\sigma_{j}} \delta_{\sigma_{0},\sigma_{k}} \nonumber \\
    &- \frac{1}{2} \sum_{j \neq k \neq l}^{2M-1} \frac{1 - z_{j}^{2}}{(z_{j} - z_{k})(z_{j} - z_{l})} (1 - \delta_{\sigma_{k},\sigma_{l}}) - \sum_{j \neq k \neq l}^{2M-1} \frac{z_{j}}{z_{j} - z_{l}} [\delta_{\sigma_{0},\sigma_{j}}(1 - \delta_{\sigma_{0},\sigma_{k}})\delta_{\sigma_{0},\sigma_{l}} + (1 - \delta_{\sigma_{0},\sigma_{j}})\delta_{\sigma_{0},\sigma_{k}}(1 - \delta_{\sigma_{0},\sigma_{l}})] \nonumber \\
    &+ \frac{1}{3}M(M-1)(M-2).
\end{align}
Making use of the identities
\begin{align}
    &\sum_{j \neq k \neq l}^{2M-1} \frac{1 - z_{j}^{2}}{(z_{j} - z_{k})(z_{j} - z_{l})} (1 - \delta_{\sigma_{k},\sigma_{l}}) \nonumber \\
    =& \sum_{j \neq k \neq l}^{2M-1} \frac{1 - z_{j}^{2}}{(z_{j} - z_{k})(z_{j} - z_{l})} \delta_{\sigma_{j},\sigma_{k}}(1 - \delta_{\sigma_{k},\sigma_{l}}) + \sum_{j \neq k \neq l}^{2M-1} \frac{1 - z_{j}^{2}}{(z_{j} - z_{k})(z_{j} - z_{l})} (1 - \delta_{\sigma_{j},\sigma_{k}})(1 - \delta_{\sigma_{k},\sigma_{l}}) \nonumber \\
    =& \sum_{j \neq k \neq l}^{2M-1} \frac{1 - z_{j}^{2}}{(z_{j} - z_{k})(z_{j} - z_{l})} (\delta_{\sigma_{j},\sigma_{k}} + \delta_{\sigma_{j},\sigma_{l}})(1 - \delta_{\sigma_{k},\sigma_{l}}) = 2 \sum_{j \neq k \neq l}^{2M-1} \frac{1 - z_{j}^{2}}{(z_{j} - z_{k})(z_{j} - z_{l})} \delta_{\sigma_{j},\sigma_{k}}(1 - \delta_{\sigma_{k},\sigma_{l}}),
\end{align}
\begin{equation}
    \sum_{j \neq k \neq l}^{2M-1} \frac{z_{j}}{z_{j} - z_{l}} \delta_{\sigma_{0},\sigma_{j}}(1 - \delta_{\sigma_{0},\sigma_{k}})\delta_{\sigma_{0},\sigma_{l}} = \frac{1}{2} M(M-1)(M-2),
\end{equation}
\begin{equation}
    \sum_{j \neq k \neq l}^{2M-1} \frac{z_{j}}{z_{j} - z_{l}} (1 - \delta_{\sigma_{0},\sigma_{j}})\delta_{\sigma_{0},\sigma_{k}}(1 - \delta_{\sigma_{0},\sigma_{l}}) = \frac{1}{2} M (M-1)^{2},
\end{equation}
\eqref{eq:auxiliary-identity-1} and~\eqref{eq:auxiliary-identity-2}, and substituting~\eqref{eq:summation-evaluated} into~\eqref{eq:HB-action}, one ends up with
\begin{align}
\label{eq:HB-action-result}
    H_{\textrm{B2}} \Psi = \Bigg[ &-\frac{3}{2} \lambda \sum_{j \neq k}^{2M-1} \frac{1 - z_{j}z_{k}}{(z_{j} - z_{k})^{2}} - 2 \lambda \sum_{j \neq k}^{2M-1} \frac{1 + z_{j}}{z_{j} - z_{k}} \delta_{\sigma_{0},\sigma_{j}} - 2 \lambda \sum_{j=1}^{2M-1} \frac{z_{j}}{1 - z_{j}} \delta_{\sigma_{0},\sigma_{j}} + 2 \lambda \sum_{j \neq k}^{2M-1} \frac{1 - z_{j}^{2}}{(z_{j} - z_{k})^{2}} \delta_{\sigma_{j},\sigma_{k}} \nonumber \\
    &+ 2 \lambda \sum_{j \neq k \neq l}^{2M-1} \frac{1 - z_{j}^{2}}{(z_{j} - z_{k})(z_{j} - z_{l})} \delta_{\sigma_{j},\sigma_{k}} + \frac{\lambda}{3}(M-1)(8M^{2}-7M-6) \Bigg] \Psi,
\end{align}
where we have made use of~\eqref{eq:auxiliary-identity-0} again.

\subsubsection{The Kondo term}

The term of Kondo interactions can similarly be rewritten in terms of permutation operators as
\begin{equation}
    H_{\textrm{I2}} = \frac{1}{6}(2\lambda + 1)\sum_{j=1}^{2M-1} \frac{1}{1 - z_{j}} ~(2\mathrm{P}_{0j} - 1),
\end{equation}
of which the action on the wave function in~\eqref{eq:wave-function-continuum} reads
\begin{equation}
    H_{\textrm{I2}} \Psi = - \frac{1}{6}(2\lambda + 1)\sum_{j=1}^{2M-1} \frac{1}{1 - z_{j}} \Psi + \frac{1}{3}(2\lambda + 1)\sum_{j=1}^{2M-1} \frac{1}{1 - z_{j}} \delta_{\sigma_{0},\sigma_{j}} \Psi + \frac{1}{3}(2\lambda + 1) \sum_{j=1}^{2M-1} \frac{1}{1 - z_{j}} (1 - \delta_{\sigma_{0},\sigma_{j}}) \mathrm{P}_{0j} \Psi.
\end{equation}
Obviously, the only non-trivial term on the right-hand-side of the above expression is the last one. Let us consider this term separately:
\begin{align}
    & \sum_{j=1}^{2M-1} \frac{1}{1 - z_{j}} (1 - \delta_{\sigma_{0},\sigma_{j}}) \mathrm{P}_{0j} \Psi = - \sum_{j=1}^{2M-1} \frac{1}{1 - z_{j}} (1 - \delta_{\sigma_{0},\sigma_{j}}) \prod_{l(\neq j)}^{2M-1} \left( \frac{z_{j} - z_{l}}{1 - z_{l}} \right)^{\delta_{\sigma_{0},\sigma_{l}} - \delta_{\sigma_{j},\sigma_{l}}} \Psi \nonumber \\
    =& - \sum_{k=1}^{M} \frac{1}{1 - v_{k}} \prod_{m=1}^{M-1} \left( \frac{v_{k} - u_{m}}{1 - u_{m}} \right) \prod_{n(\neq k)}^{M} \left( \frac{1 - v_{n}}{v_{k} - v_{n}} \right) \Psi.
\end{align}
Again, this summation can be calculated by applying the technique based on complex analysis. This time, the auxiliary function is chosen to be
\begin{equation}
    h(z) = \frac{1}{(1 - z)^{2}} \prod_{m=1}^{M-1} \left( \frac{z - u_{m}}{1 - u_{m}} \right) \prod_{n=1}^{M} \left( \frac{1 - v_{n}}{z - v_{n}} \right).
\end{equation}
At poles $z = v_{k}, k = 1, \ldots, M$ and $z = 1$, the residues of $h(z)$ are
\begin{equation}
    \mathrm{Res}(h,v_{k}) = \frac{1}{1 - v_{k}} \prod_{m=1}^{M-1} \left( \frac{v_{k} - u_{m}}{1 - u_{m}} \right) \prod_{n(\neq k)}^{M} \left( \frac{1 - v_{n}}{v_{k} - v_{n}} \right)
\end{equation}
and
\begin{equation}
    \mathrm{Res}(h,1) = \sum_{j=1}^{M-1} \frac{1}{1 - u_{j}} - \sum_{k=1}^{M} \frac{1}{1 - v_{k}},
\end{equation}
respectively, and $\mathrm{Res}(h,\infty) = 0$. According to Cauchy's theorem, one immediately finds
\begin{equation}
    \sum_{k=1}^{M} \frac{1}{1 - v_{k}} \prod_{m=1}^{M-1} \left( \frac{v_{k} - u_{m}}{1 - u_{m}} \right) \prod_{n(\neq k)}^{M} \left( \frac{1 - v_{n}}{v_{k} - v_{n}} \right) = - \sum_{j=1}^{M-1} \frac{1}{1 - u_{j}} + \sum_{k=1}^{M} \frac{1}{1 - v_{k}}.
\end{equation}
Hence, for the action of the Kondo term on $\Psi$, we have
\begin{equation}
\label{eq:HK-action-result}
    H_{\textrm{I2}} \Psi = -\frac{1}{2} (2\lambda + 1) \sum_{j=1}^{2M-1} \frac{1}{1 - z_{j}} (1 - 2\delta_{\sigma_{0},\sigma_{j}}) \Psi.
\end{equation}

\subsubsection{The terms of two-body potential}

Summarizing the actions of the above three parts of the Hamiltonian on $\Psi$ [\eqref{eq:H0-action-result},~\eqref{eq:HB-action-result} and~\eqref{eq:HK-action-result}], we have
\begin{align}
    (H_{0} + H_{\textrm{B2}} + H_{\textrm{I2}}) \Psi = \Bigg[ & - \lambda (\lambda + \frac{1}{2}) \sum_{j \neq k}^{2M-1} \frac{1 - z_{j}z_{k}}{(z_{j} - z_{k})^{2}} - (\lambda + \frac{1}{2}) \sum_{j=1}^{2M-1} \frac{1}{1 - z_{j}}  \nonumber \\
    & + \frac{\lambda^{2}}{3} (M-1)(2M-1)(4M-3) + \frac{\lambda}{3} M(M-1)(8M-7) + \frac{1}{3} M(M-1)(2M-1) \Bigg] \Psi,
\end{align}
where the coordinate-dependent terms are precisely cancelled by the action of the term of two-body potentials in $H$,
\begin{equation}
    H_{\textrm{B1}} + H_{\textrm{I1}} = (\lambda + \frac{1}{2}) \sum_{j=1}^{2M-1} \frac{1}{1 - z_{j}} + \lambda (\lambda + \frac{1}{2}) \sum_{j \neq k}^{2M-1} \frac{1 - z_{j}z_{k}}{(z_{j} - z_{k})^{2}}.
\end{equation}
Thus, we conclude that the wave function in~\eqref{eq:wave-function-continuum} is an exact eigenstate of the total Hamiltonian, $H \Psi = E \Psi$, where the energy eigenvalue is precisely given by~\eqref{eq:eigenenergy}.

\subsection{Transcribing the kinetic term to the lattice}

In this last subsection of Sec.~\ref{sec:parent-hamiltonian}, we show that the lattice counterpart of the kinetic term in the continuum space,
\begin{equation}
    H_{0} = -\sum_{j=1}^{2M-1} \frac{\partial^{2}}{\partial x_{j}^{2}},
\end{equation}
is indeed given by the hopping Hamiltonian defined as
\begin{align}
    \widetilde{H}_{0} = \sum_{j=1}^{N} \left[ \frac{1}{3}N^{2} + \frac{1}{6} - \frac{1}{2(1-w_{j}^2)} \right] n_{j} + 2 \sum_{j \neq k}^{N} \sum_{\sigma} (-1)^{j-k} \left[ \frac{1}{\left( d(\theta_{j},\theta_{k}) \right)^{2}} - \frac{1}{\left( \bar{d}(\theta_{j},\theta_{k}) \right)^{2}} \right] c_{w_{k},\sigma}^{\dagger}c_{w_{j},\sigma},
\end{align}
where $\theta_{j},~j = 1, \ldots, N$ are angular coordinates of the sites on which the itinerant particles can hop, $w_{j} \equiv \mathrm{cos}~\theta_{j}$. To this end, it suffices to prove that when acting on arbitrary single-particle wave function, the effect of $\widetilde{H}_{0}$ is the same as that of $H_{0}$. With the choice $\theta_{j} = \pi (j - 1/2)/N,~j = 1, \ldots, N$, a complete single-particle basis is given by $\{ T_{q}(w_{j}) \}_{q=0}^{N-1}$, where
\begin{equation}
    T_{q}(z) = T_{q}(\mathrm{cos}x) = \mathrm{cos}(qx)
\end{equation}
are the~\emph{Chebyshev polynomials of the first kind}. The action of the hopping term on this basis reads
\begin{align}
    &\left[ \frac{1}{3}N^{2} + \frac{1}{6} - \frac{1}{2(1-w_{j}^2)} \right] T_{q}(w_{j}) + 2 \sum_{k (\neq j)}^{N} (-1)^{j-k} \left[ \frac{1}{\left( d(\theta_{j},\theta_{k}) \right)^{2}} - \frac{1}{\left( \bar{d}(\theta_{j},\theta_{k}) \right)^{2}} \right] T_{q}(w_{k}) \nonumber \\
    =~& \frac{2}{N} \sum_{k=1}^{N} \sum_{p=0}^{N-1} p^{2} T_{p}(w_{j})T_{p}(w_{k})T_{q}(w_{k}) \nonumber \\
    =~& q^{2} T_{q}(w_{j}) = - \frac{\mathrm{d}^{2}}{\mathrm{d}x^{2}} T_{q}(z)\vert_{z = w_{j}},
\end{align}
where we have made use of the summation identity~\cite{tu2019}
\begin{equation}
    \sum_{q=1}^{N-1} q^{2} \mathrm{cos}\left( \frac{q\pi}{N}l \right) = \begin{cases}
    \begin{array}{c}
    \frac{1}{2} (-1)^{l} N \left[ \mathrm{sin}^{-2}\left( \frac{\pi}{2N}l \right) - N \right] \\
    \frac{1}{6}N(N-1)(2N-1) \\
    \end{array} & \begin{array}{c}
    l = 1, \ldots, 2N-1 \\
    l = 0
    \end{array} \end{cases}
\end{equation}
and the discrete orthogonality of Chebyshev polynomials,
\begin{equation}
    \sum_{k=1}^{N} T_{p}(w_{k})T_{q}(w_{k}) = \frac{N}{2}\delta_{pq},\quad q \neq 0.
\end{equation}
Indeed, acting the hopping term on the single-particle basis amounts to taking second-order derivative with respect to the angular coordinate $x$, which is precisely the same as the effect of acting the kinetic term. Thus, by replacing $H_{0}$ with $\widetilde{H}_{0}$, the lattice version of the wave function in~\eqref{eq:wave-function-continuum} is an exact eigenstate of the total Hamiltonian, with eigenvalue still given by~\eqref{eq:eigenenergy}.

\section{Model wave functions as CFT correlators}

This section aims to prove that the model wave functions on the lattice can be expressed as certain chiral correlators in the free boson CFT.

The chiral bosonic field $\phi(w)$ is defined in terms of its mode expansion~\cite{francesco1997}:
\begin{equation}
    \phi(w) = Q - iP~\mathrm{ln}w + i\sum_{k \neq 0}\frac{1}{k}a_{k}w^{-k},
\end{equation}
in which the modes satisfy the familiar commutation relations,
\begin{equation}
\label{eq:bosonic-algebra}
    [Q,P] = i,\quad [a_{k},a_{k^{\prime}}] = k\delta_{k+k^{\prime}}.
\end{equation}
$\phi(w)$ can be used to construct the vertex operator
\begin{equation}
    :e^{i\phi(w)}: = \mathrm{exp}\left( iQ + \sum_{k > 0}\frac{1}{k}a_{-k}w^{k} \right) \mathrm{exp}\left( P~\mathrm{ln}w - \sum_{k > 0}\frac{1}{k}a_{k}w^{-k} \right),
\end{equation}
where $:\cdots:$ denotes normal ordering, which amounts to moving all operators $a_{k>0}$ and $P$ (which annihilate the vacuum $\vert 0 \rangle$,~$a_{k>0} \vert 0 \rangle = P \vert 0 \rangle = 0$) to the right of other operators in ``$\cdots$''. Making use of~\eqref{eq:bosonic-algebra}, it is straightforward to show that
\begin{align}
\label{eq:vertex-correlator-plane}
    :e^{i\alpha_{1}\phi(w_{1})}: \cdots :e^{i\alpha_{N}\phi(w_{N})}: = &~\mathrm{exp}\left( i \sum_{j=1}^{N} \alpha_{j} Q + \sum_{j=1}^{N} \sum_{k > 0} \frac{\alpha_{j}}{k}a_{-k}w_{j}^{k} \right) \mathrm{exp}\left( \sum_{j=1}^{N} \left( \alpha_{j}\mathrm{ln}w_{j} \right)P - \sum_{j=1}^{N} \sum_{k > 0}\frac{\alpha_{j}}{k}a_{k}w_{j}^{-k} \right) \nonumber \\
    &\times \prod_{1 \leq j<j^{\prime} \leq N} \left( w_{j} - w_{j^{\prime}} \right)^{\alpha_{j}\alpha_{j^{\prime}}}.
\end{align}
In the main text, the vertex operators being used to construct model wave functions are composed of a bosonic field with two components $\phi_{\textrm{s}}$ and $\phi_{\textrm{c}}$:
\begin{align}
    A_{0}^{\uparrow/\downarrow}(1) &= \kappa_{\uparrow/\downarrow} :e^{\pm \frac{i}{\sqrt{2}}\phi_{\textrm{s}}(1)} e^{\frac{i}{\sqrt{2(2\lambda + 1)}}\phi_{\textrm{c}}(1)}:, \\
    A^{\uparrow}(u_{k}) &= \kappa_{\uparrow} :e^{\frac{i}{\sqrt{2}}\phi_{\textrm{s}}(u_{k})} e^{i\sqrt{\frac{2\lambda + 1}{2}}\phi_{\textrm{c}}(u_{k})}:, \\
    A^{\downarrow}(v_{m}) &= \kappa_{\downarrow} :e^{-\frac{i}{\sqrt{2}}\phi_{\textrm{s}}(v_{m})} e^{i\sqrt{\frac{2\lambda + 1}{2}}\phi_{\textrm{c}}(v_{m})}:,
\end{align}
where $\kappa_{\uparrow/\downarrow}$ are Klein factors that satisfy the algebra of Majorana fermions,
\begin{equation}
    \kappa_{\uparrow}^{2} = \kappa_{\downarrow}^{2} = 1, \quad \{ \kappa_{\uparrow},\kappa_{\downarrow} \} = 0.
\end{equation}
Using~\eqref{eq:vertex-correlator-plane} for both $\phi_{\textrm{s}}$ and $\phi_{\textrm{c}}$, taking care of the signs resulting from permuting the Klein factors and introducing a ``neutralizing background charge''
\begin{equation}
    \mathcal{O}_{\textrm{bg}} = \mathrm{exp}\left[ -i\left( \frac{1}{\sqrt{2(2\lambda + 1)}} + (2M-1)\sqrt{\frac{2\lambda + 1}{2}} \right) Q_{\textrm{c}} \right]
\end{equation}
to compensate the zero-mode part in~\eqref{eq:vertex-correlator-plane}, one finds that the model wave functions are equivalent to chiral correlators of these vertex operators evaluated in the CFT vacuum $\vert 0 \rangle$:
\begin{align}
    &\langle \mathcal{O}_{\textrm{bg}} A_{0}^{\uparrow}(1) A^{\uparrow}(u_{1}) \cdots A^{\uparrow}(u_{M-1}) A^{\downarrow}(v_{1}) \cdots A^{\downarrow}(v_{M}) \rangle \nonumber \\
    =&~\left( \prod_{k=1}^{M-1} (1 - u_{k}) \right) \left( \prod_{1 \leq k < l \leq M-1} (u_{k} - u_{l})^{\lambda+1} \prod_{1 \leq m < n \leq M} (v_{m} - v_{n})^{\lambda+1} \right) \left( \prod_{k=1}^{M-1} \prod_{m=1}^{M} (u_{k} - v_{m})^{\lambda} \right) \nonumber \\
    =&~\langle \Uparrow; u_{1}, \ldots, u_{M-1}, v_{1}, \ldots, v_{M} \vert \widetilde{\Psi} \rangle,
\end{align}
\begin{align}
    &\langle \mathcal{O}_{\textrm{bg}} A_{0}^{\downarrow}(1) A^{\uparrow}(u_{1}) \cdots A^{\uparrow}(u_{M}) A^{\downarrow}(v_{1}) \cdots A^{\downarrow}(v_{M-1}) \rangle \nonumber \\
    =&~(-1)^{M} \left( \prod_{m=1}^{M-1} (1 - v_{m}) \right) \left( \prod_{1 \leq k < l \leq M} (u_{k} - u_{l})^{\lambda+1} \prod_{1 \leq m < n \leq M-1} (v_{m} - v_{n})^{\lambda+1} \right) \left( \prod_{k=1}^{M} \prod_{m=1}^{M-1} (u_{k} - v_{m})^{\lambda} \right) \nonumber \\
    =&~\langle \Downarrow; u_{1}, \ldots, u_{M}, v_{1}, \ldots, v_{M-1} \vert \widetilde{\Psi} \rangle,
\end{align}
where we have chosen the sign convention $\kappa_{\uparrow}\kappa_{\downarrow} = +1$.

\section{Low-energy spectrum}
\label{sec:spectrum}

\begin{figure}
\includegraphics[width=0.50\columnwidth]{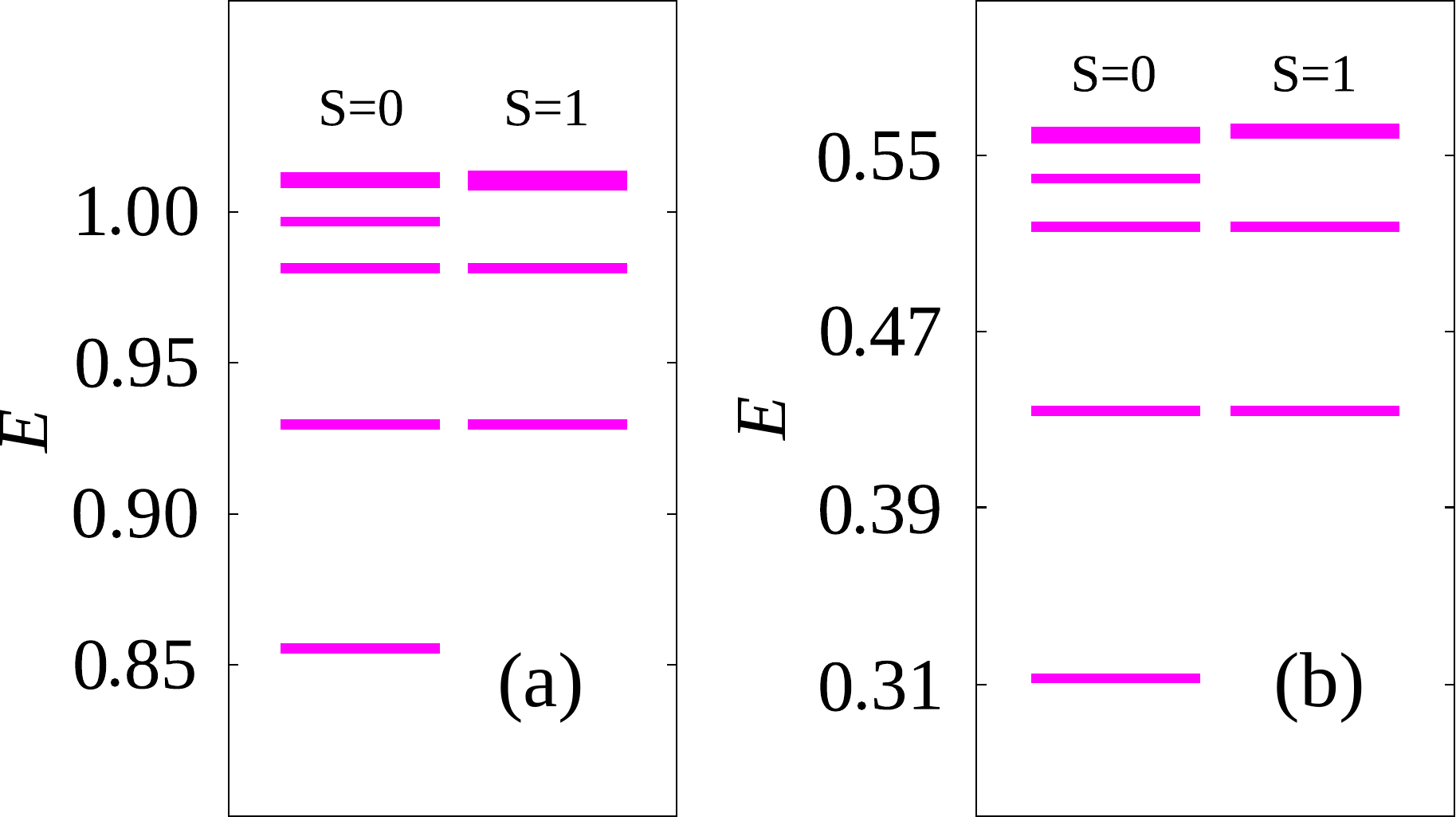}
\caption{Energy spectra of the Hamiltonian $\widetilde{H}/N^{2}$ on lattice. The total spin $S$ of the levels are indicated. (a) $\lambda=1$, $N=16$, and $M=4$; (b) $\lambda=2$, $N=16$, and $M=4$.}
\label{fig:spectrum}
\end{figure}

In this section, we present the low-lying energy levels of the lattice model. Each level represents a multiplet due to the SU(2) spin symmetry. For computational ease, the Hamiltonian has been rescaled by a factor $1/N^{2}$. This causes the apparent closeness of some eigenvalues in Fig.~\ref{fig:spectrum}, but they are clearly different that can be resolved easily in Lanczos sparse matrix diagonalization. As an example, the difference between the first and second excited states in Fig.~\ref{fig:spectrum}(a) is about $0.0003$. This suggests that the lattice models do \emph{not} have Yangian-type conserved quantities, at least not exact ones.

\section{Out-of-time-ordered correlator}
\label{sec:otoc}

\begin{figure}
\includegraphics[width=0.65\columnwidth]{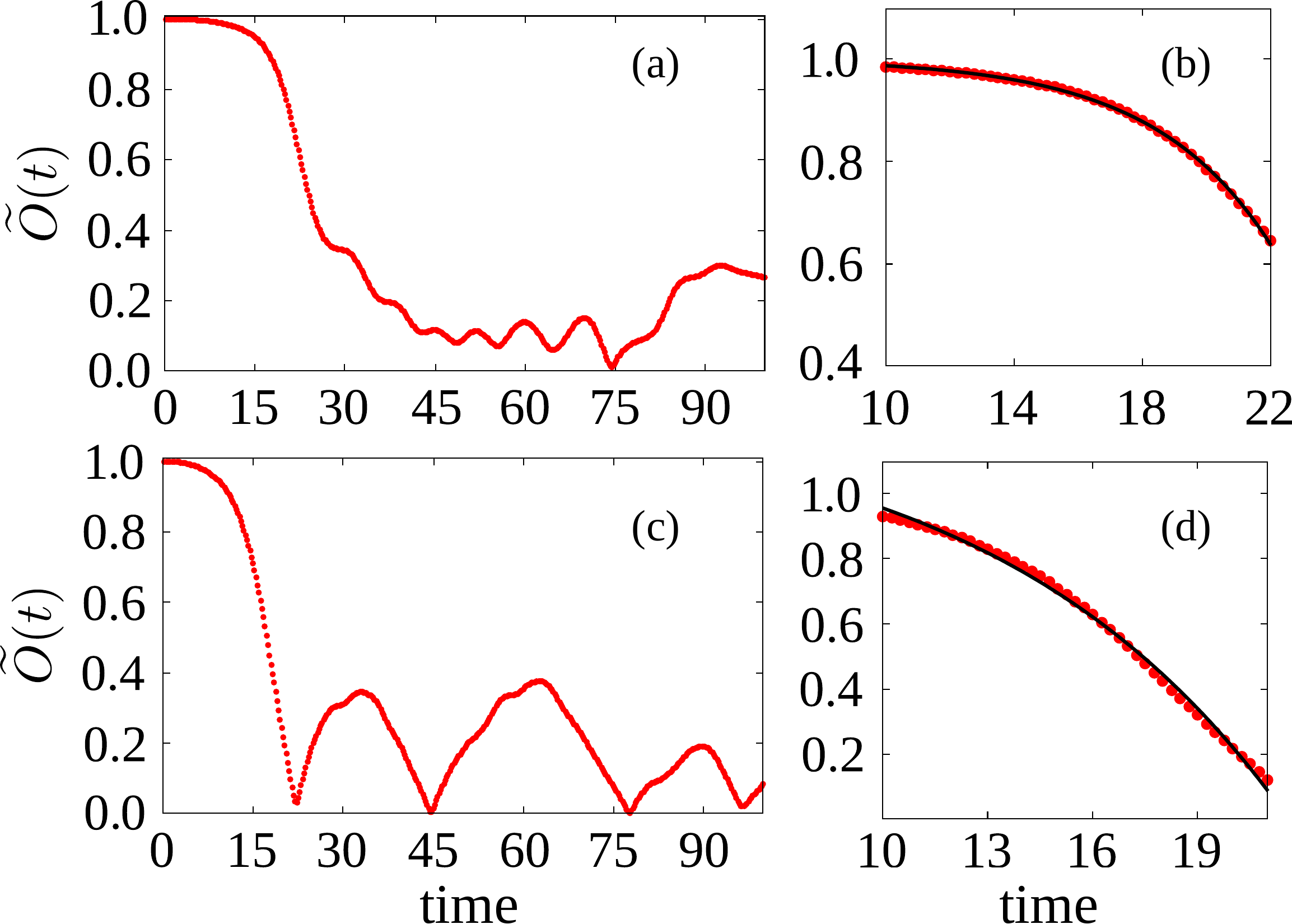}
\caption{Out-of-time-ordered correlator of the lattice model. (a) $\lambda=1$, $N=12$, and $M=3$. (b) The data points in the interval $[10,22]$ of panel (a) are fitted using the curve $1.0014-9.36\exp(0.2712t)/10^{4}$. (c) $\lambda=2$, $N=12$, and $M=3$. (d) The data points in the interval $[10,21]$ of panel (c) are fitted using the curve $1.2764-0.0972\exp(0.1193t)$.}
\label{fig:otoc}
\end{figure}

For two local operators $\widehat{W}(t)$ and $\widehat{V}(t)$ in the Heisenberg picture, their out-of-time-ordered correlator (OTOC) is
\begin{eqnarray}
O(t) = \langle \widehat{W}^{\dag}(t) \widehat{V}^{\dag}(0) \widehat{W}(t) \widehat{V}(0) \rangle,
\end{eqnarray}
where the expectation value may be computed with respect to a pure state or a thermal state at a specific temperature~\cite{Larkin1969,Maldacena2016}. We shall study the OTOC of the ground state $|\widetilde{\Psi}\rangle$ given by Eq.~(5) of the main text. The time evolution of operators is defined using the parent Hamiltonian $\widetilde{H}$ associated with $|\widetilde{\Psi}\rangle$. The correlator can be rewritten as $O(t) = \langle \Phi_{B} | \Phi_{A} \rangle$ with
\begin{eqnarray}
| \Phi_{A} \rangle = \exp(i\widetilde{H}t) \widehat{W}(0) \exp(-i\widetilde{H}t) \widehat{V}(0) | \widetilde{\Psi} \rangle
\end{eqnarray}
and
\begin{eqnarray}
| \Phi_{B} \rangle = \widehat{V}(0) \exp(i\widetilde{H}t) \widehat{W}(0) \exp(-i\widetilde{H}t) | \widetilde{\Psi} \rangle.
\end{eqnarray}
The OTOC is further normalized to be
\begin{eqnarray}
\widetilde{O}(t) = \frac{\langle \Phi_{B} | \Phi_{A} \rangle}{\sqrt{\langle \Phi_{A} | \Phi_{A} \rangle \langle \Phi_{B} | \Phi_{B} \rangle}}.
\end{eqnarray}
It has been proposed that the OTOC is a useful diagnostic of quantum chaos. To be specific, we have
\begin{eqnarray}
\widetilde{O}(t) = f_{0} - f_{1} \exp\left[ \lambda_{L}(t-t_{0}) \right]
\label{eq:expdecay}
\end{eqnarray}
with $f_{0}{\approx}1$ if the system is chaotic. This is a manifestation of the butterfly effect because $\widetilde{O}(t)$ deviates from unity in an exponential manner starting at time $t_{0}$. The coefficient $\lambda_{L}$ is called the Lyapunov exponent.

In our calculations, $\widehat{V}$ is the $z$-component spin on the impurity site and $\widehat{W}$ is the $z$-component spin on the leftmost lattice site for the itinerant particles. As one can see from Fig.~\ref{fig:otoc}, $\widetilde{O}(t)$ stays close to unity at first, then decays quickly for some time, and finally oscillates with small amplitudes. It seems reasonable to claim that OTOC also indicates that the model is non-integrable. The OTOC has been fitted using Eq.~\eqref{eq:expdecay} in a small time window and the match is quite good in view of the limited system sizes accessed by exact diagonalization.


\end{document}